\documentclass{emulateapj}
\usepackage{amsmath}

\shorttitle{Dynamical mass of a $z \sim 1.3$ QSO host galaxy}
\shortauthors{Inskip et al.}

\begin{document}

\title{Resolving the dynamical mass of a $z \sim 1.3$ QSO host galaxy using SINFONI and Laser Guide Star assisted Adaptive Optics. }

\author{K. J. Inskip, K. Jahnke, H.-W. Rix \& G. van de Ven}
\affil{Max Planck Institute for Astronomy, K\"onigstuhl 17, 69117 Heidelberg, Germany}
\email{inskip@mpia.de}

\begin{abstract}
Recent studies of the tight scaling relations between the masses of supermassive black holes and their host galaxies have suggested that in the past black holes constituted a larger fraction of their host galaxies' mass. However, these arguments are limited by selection effects and difficulties in determining robust host galaxy masses at high redshifts.
Here we report the first results of a new, complementary diagnostic route: we directly determine a dynamical host galaxy mass for the $z=1.3$ luminous quasar J090543.56+043347.3 through high-spatial-resolution ($0.47^{\prime\prime}$, 4\,kpc FWHM) observations of the host galaxy gas kinematics  over 30$\times$40 kpc using ESO/VLT/SINFONI with LGS/AO. Combining our result of $\rm M_{dyn} = 2.05^{+1.68}_{-0.74} \times 10^{11}\, \rm M_{\odot}$ (within a radius $5.25 \pm  1.05$\,kpc) with $\rm M_{BH, MgII} = 9.02 \pm 1.43 \times 10^8\, \rm M_{\odot}$,  $\rm M_{BH, H\alpha} = 2.83^{+1.93}_{-1.13} \times 10^8\, \rm M_{\odot}$, we find that the ratio of  black hole mass to host galaxy dynamical mass for J090543.56+043347.3 matches the present-day relation for $\rm M_{BH}$ vs.\ $\rm M_{Bulge,Dyn}$, well within the IR scatter, deviating at most a factor of two from the mean.  J090543.56+043347.3  displays clear signs of an ongoing tidal interaction and of spatially extended star formation at a rate of $50-100\, \rm M_{\odot}\, \rm yr^{-1}$, above the cosmic average for a galaxy of this mass and redshift. We argue that its subsequent evolution may move J090543.56+043347.3 even closer to the $z=0$ relation for $\rm M_{BH}$ vs.\ $\rm M_{Bulge,Dyn}$.
Our results support the picture where any substantive evolution in these relations must occur prior to $z \sim 1.3$. Having demonstrated the power of this modelling approach we are currently analyzing similar data on seven further objects to better constrain such evolution.

\end{abstract}


\section{Introduction}

The tight correlations between the properties of supermassive black holes (BHs) and their host galaxies \citep{kr95, mag98, geb00, fm00, tre02, md02, mh03, hr04, grah04, kb09, g09, jah09, mer10, grah11} are a powerful tool for probing their relative and possibly interdependent evolution and growth across cosmic time.  Since the hierarchical assembly in conjunction with the central limit theorem can explain the existence of these correlations \citep{peng07, hirs10, jahn11}, evolution in the correlations may solely be due to the relative growth speeds of the black hole mass and the corresponding total and bulge stellar masses of the host galaxy as a function of redshift. Hence, evolution in these correlations is a key diagnostic of the growth channels and fuelling mechanisms of black holes, active galactic nuclei (AGN) and their host galaxies.

Overall, the current picture is one of evolution in the various correlations clearly being strong at $z > 2$ \citep{wal04, peng06a, peng06b, som09, deca10b}. A number of studies also find a strong evolution signal at low redshifts \citep{woo06, treu07, benn10}, particularly in terms of correlations involving the bulge properties of the host galaxies: it is generally recognised that black holes of a given mass are hosted by increasingly less massive spheroids/bulges at larger values of $z$. However, at lower redshifts the majority of studies show a mild (by factor $\la$2-3) or absent evolution, particularly where the total stellar mass of the host galaxy is concerned. For example, for AGN at $1 \la z \la 2$ \citet{mer10} observe an offset of a factor of 2-3 from the $z=0$ scaling relation, and note that given the large scatter, the data could still be consistent with zero evolution at the 2$\sigma$ level. And while \citet{deca10b} observe an overall evolution of up to a factor of 7 between $z=0$ and $z=3$, the sources observed at intermediate ($1.0 < z < 1.5$) and low ($z<1$) redshifts are consistent with modest (0.3dex) and no evolution respectively.

Beyond the nearby Universe, observational estimates of black hole and galaxy masses are subject to considerable uncertainties. Furthermore, different measures of galaxy mass (i.e. bulge mass, stellar mass, total mass) are frequently used, necessitating additional care in the interpretation of results from different studies.  The major uncertainties of previous work in this field affect both large and small samples equally, and arise from the accuracy with which the masses of black hole and host galaxy (total and/or bulge mass) can be determined. Outside the local universe, black hole masses can only realistically be obtained for currently accreting type-1 AGN (i.e. quasars or Seyfert-1 galaxies). For such objects, $\rm M_{BH}$ can be constrained from the combination of the UV-luminosity and the broad emission line widths from ionization models which assume virialised motion and are in turn calibrated via reverberation mapping on local AGN with a statistical uncertainty within a factor of 3-4 for Mg\textsc{ii} and C\textsc{iv} \citep{vest02, mj02, shie03, gh05, vp06, woo06, koll06, salv07, nt07, treu07, mcg08, ok08}, or less for Balmer lines  \citep{denn08, sk10}, plus any further systematic uncertainty associated with the calibration between reverberation mapping BH masses and dynamical BH masses at z=0.

However, obtaining accurate (total and/or bulge) dynamical masses for the quasar host galaxy, the second key ingredient of the correlation in question, is very difficult for objects beyond $z>0.5$.  The presence of an active nucleus makes stellar velocity dispersion measurements from absorption lines infeasible, and the estimated  velocity dispersions derived from narrow emission line widths \citep{shie03, shie06, salv07, ho08, riec08, riec09} are fraught with uncertainty \citep{ho07, gre09}.   Inferring stellar masses from imaging-based measurements of the host galaxy stellar luminosity is another possible approach used by a number of previous studies \citep{bory05, peng06a, peng06b, jah09, som09, mer10} but, as the mass-to-light ratios deduced from the UV/optical spectral energy distributions of star-forming galaxies are severely model-dependent, inferred stellar masses are non-unique/uncertain.  At redshifts $z >4$, practical mass determination is limited to methods such as the determination of dynamical masses via CO observations \citep[e.g.][]{wal04}. Direct measurement of a host galaxy's dynamical mass is free of many of the problems associated with other methods, and can equally be applied to observations at lower redshifts.  Aside from mm-observations of CO emission, dynamical masses can also be measured from ISM emission lines in the rest-frame optical.

At redshifts of $z>1$, H$\alpha$ is the most prevalent powerful emission line that can be used in practice as a kinematic tracer in AGN host galaxies. Even so, successfully probing the geometry and kinematics of the H$\alpha$-emitting gas requires integral field near-infrared (near-IR) spectroscopic observations with a spatial resolution of not more than a few kpc (i.e. $\lesssim 0.5^{\prime\prime}$). A sufficiently high (narrow line) luminosity is needed, but the H$\alpha$ luminosity of galaxies undergoing star formation at the average rate is easily sufficient -- along with the presence of favourably orientated and ordered velocity structures (also likely for a significant proportion of objects).  
As a pilot study for direct dynamical host galaxy masses for $z > 1$ quasars we present and model near-IR integral field spectroscopy of the z=1.3118 quasar J090543.56+043347.3, using SINFONI \citep{bon04, eis03}. We intend to present the results of similar observations of our full sample in a subsequent paper; interim analysis suggests that the required flux, orientation and ordered velocity field constraints are met for approximately 2/3 of the objects targeted.
A standard cosmological model with $\Omega = 0.27$, $\Lambda = 0.73$ and $H_0 = 71\, \rm km s^{-1} Mpc^{-1}$ is assumed throughout; this leads to a spatial scale of $8.45\, \rm kpc/arcsec$ at this redshift.

\section{Observations and data reduction}

Our target, the $z=1.31177$ SDSS quasar J090543.56+043347.3 \citep{schnei07, shen08}, was selected on the basis of the presence of a nearby guide star (12th magnitude at a distance $\sim$4.3 arcmin) to be used for tip-tilt correction. This object was observed using the near-infrared integral field spectrometer SINFONI on the Very Large Telescope of the European Southern Observatory \citep{bon04, eis03} and the PARSEC Laser Guide Star Adaptive Optics, on the night of 2009 December 15, with an average airmass of $\sim$1.16 and IFU image-seeing of 0.47arcsec FWHM. 
Six pointings of 600s were obtained, in two repeating patterns of target-sky-target in order to facilitate the removal of sky background emission, with a sky-frame declination offset of 30 arcsec from the target coordinates. The H-band filter was used with a grating central wavelength of 1650nm, giving a spectral coverage of 1.45-1.83$\mu$m and a spectral resolution of 3000. The spatial scale of each of the 32 SINFONI slitlets was 100mas; these are imaged onto 64 detector pixels apiece, giving a field of view of $3^{\prime\prime} \times 3^{\prime\prime}$ for each pointing. 

\begin{figure*}
\includegraphics[scale=0.37]{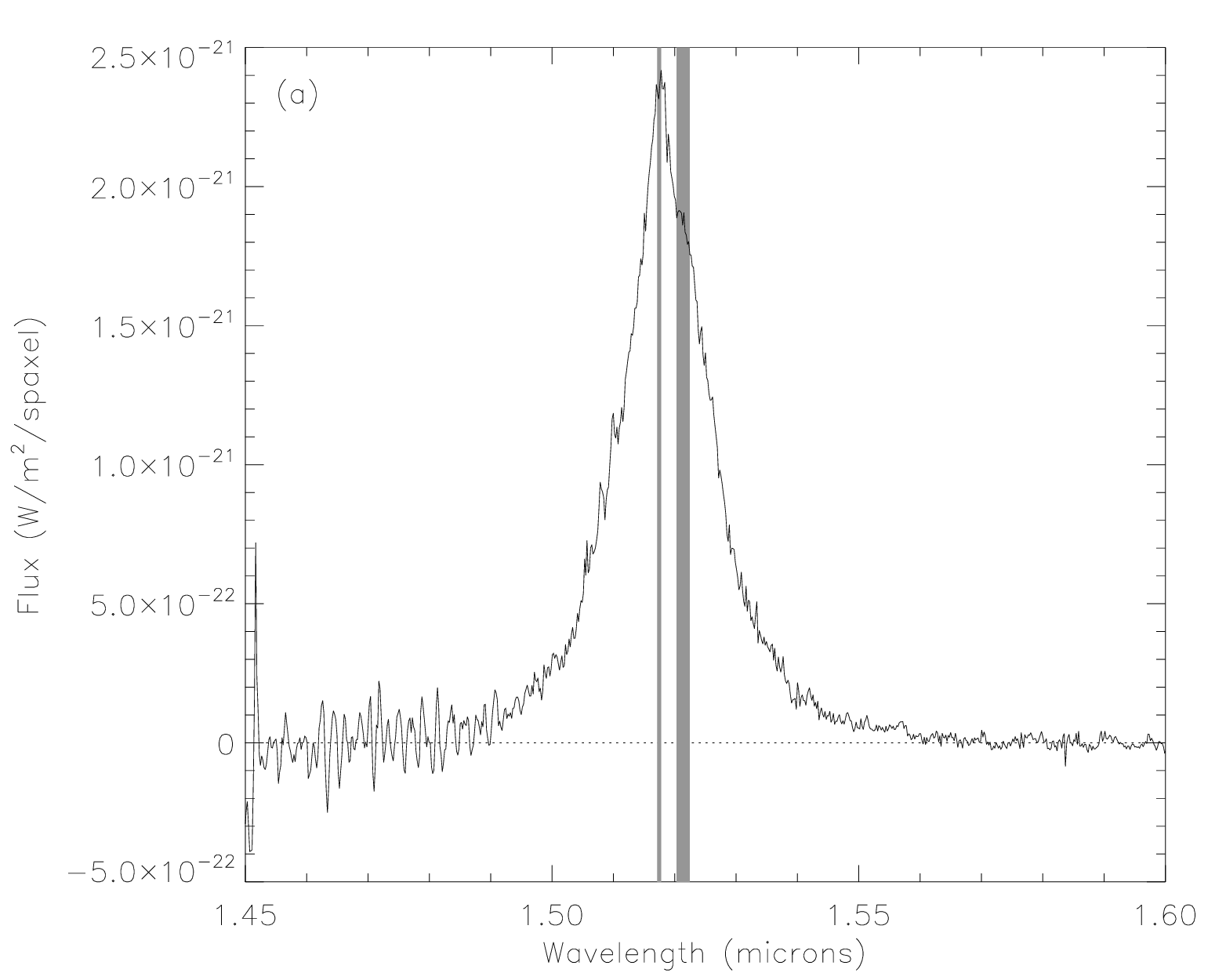}
\includegraphics[scale=0.37]{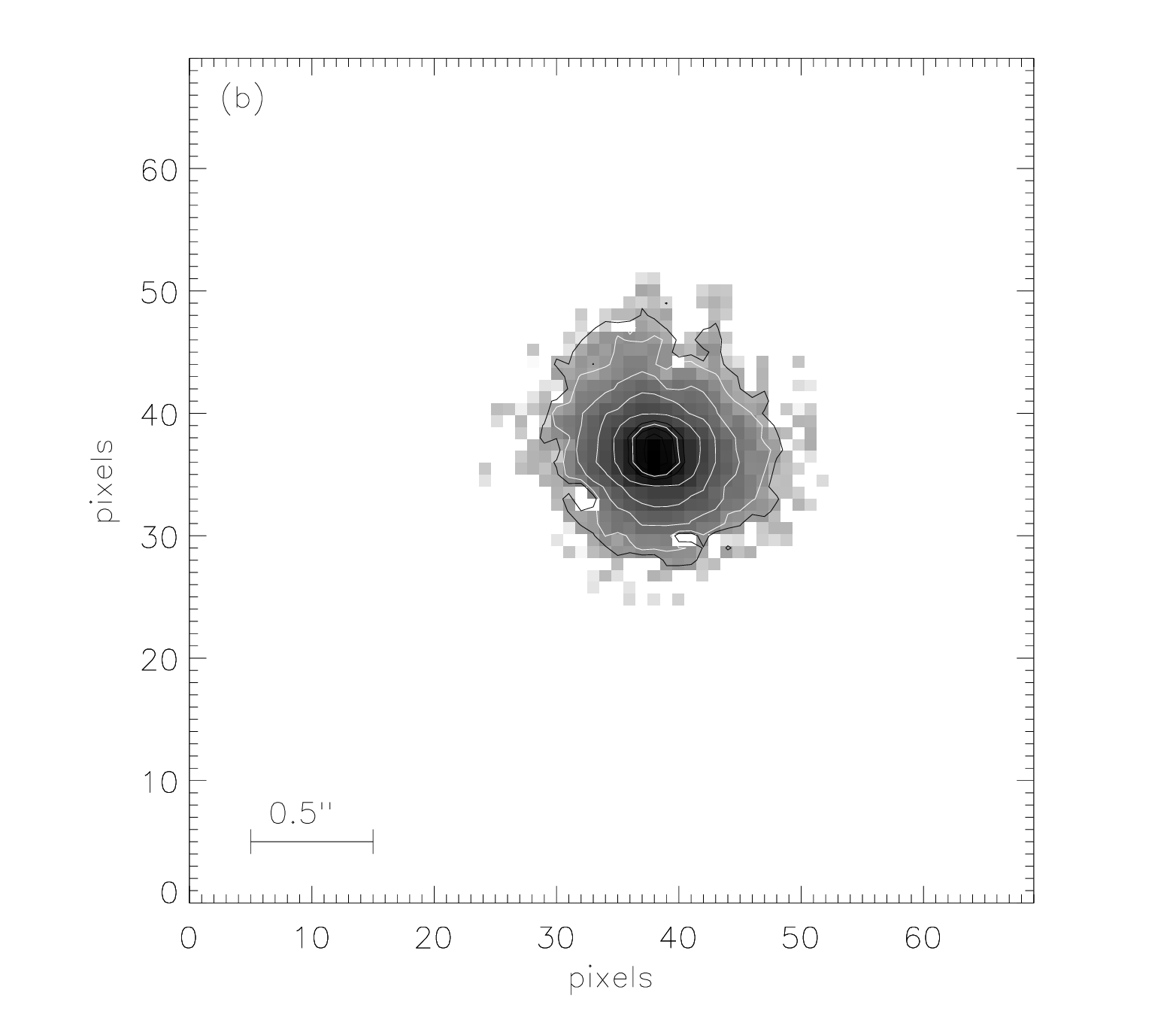}
\caption{\label{fig1} From datacube to velocity field. (a - left) Illustration of the continuum-subtracted QSO spectrum at the peak spaxel. The two wavelength regions of the broad H$\alpha$ line found to be optimal for our QSO PSF construction are highlighted in grey. They are free of contamination from narrow H$\alpha$ or [N\textsc{ii}] line emission, whether it be due to the AGN itself or the spatially extended emission from the host galaxy, or the night OH sky lines. While there are 3-4 other possible wavelength regions within the broad H$\alpha$ line which are uncontaminated by narrow line emission or night sky lines, the flux in these regions has lower signal-to-noise, and results in an overall worsening in quality of the resulting PSF profile. (b - right) The resulting PSF image defined  as the spaxel by spaxel intensity map in these three wavelength regions.   Contours are at 1, 2, 5, 10, 20 and 50\% of the peak QSO flux, with the lowest (1\%) contour in black. The 50\% and 70\% encircled energy radii are $0.12^{\prime\prime}$ and $0.2^{\prime\prime}$ respectively, corresponding to contours at 40\% and 14\% of the peak flux. 
}
\end{figure*}

Analysis of SINFONI data requires a painstaking treatment of cosmic ray hits and bad pixels, neither of which can be adequately managed by the existing pipeline reduction routines alone. We  use IRAF routines to identify bad pixels in each raw frame of data prior to applying the instrument pipeline, creating a mask file for each individual raw frame.  Otherwise, the initial $\sim$2-3\% bad pixels flagged as NaN in each raw frame would breed disproportionately as the raw data are straightened, stretched, shifted and combined within the pipeline, resulting in an unrealistic fraction of the final datacube containing NaN values (typically $\sim$30-50\%).  Ideally, we do not seek to remove signal from any pixel with some level of contamination by bad data, but rather to trace the propagation of bad pixels and thus gauge the reliability of each data element within the final datacube.   From this point onwards we work with two different data sets. Firstly, we use our final bad pixel mask to linearly interpolate over all flagged bad pixels in the observed data. Although these flagged pixels will contribute flux to their neighbours, they will only do so with locally realistic count values. Hence, we do not invent data while preserving the integrity of the spaxel grid. Secondly, mask frame images are generated and treated in the exact same manner as the observed data, thus allowing us to keep track of the behaviour and growth of the bad pixels as the datacube is generated. 

A further issue is  the mis-location of slitlet edges by the ESO SINFONI data reduction pipeline. To counter this, we apply small pixel shifts to the supplied WAVE,MAP calibration images, recovering 2-3 of the 3-4 columns of data on each side of the field of view i.e. 10\% of the field, which would otherwise be lost. The outermost columns of each slitlet which cannot be reliably wavelength calibrated remain masked out. Each raw frame and its associated mask are then {\it individually} converted from a 2-d image into 3-d datacubes using the 'sinfo-rec-jitter' routine in the pipeline and the improved skyline subtraction IDL routines of \citet{dav07}. Our bad pixel mask datacubes contain the proportional contribution of bad pixels to each spaxel in the cube. For spaxels where the proportional contribution of bad pixels is larger than 50\%, that particular spaxel in the corresponding observational data cube is converted to a value of NaN; such spaxels do not contain reliable information. This accounts for a total of approximately 6\% of spaxels in our final combined datacube.   The remaining spaxels in the datacube are either completely uncontaminated by the presence of nearby bad pixels (true for $\sim$50-70\% of spaxels) or are dominated by emission from clean spaxels ($\sim$25-45\%). As described earlier, in this latter case the spaxel flux is derived from an initial image where the flagged bad pixels/cosmic ray hits have been replaced by linear interpolation over the surrounding clean data. As these spaxels originate from a majority of good data pixels, with a minority contribution from proportionately scaled locally plausible flux values, they are judged reliable, and left as they are. 
 This done, the individual datacubes are straightened and aligned via a process of gaussian centroiding, sub-pixel linear interpolation and shifting of each wavelength slice.

Our final datacube is produced by mean-combining the individual aligned cubes on a spaxel-by-spaxel basis, ignoring masked NaN values and 3-sigma clipping any value which deviates significantly from the typical values found at that spaxel position and those of the neighbouring spaxels across all individual aligned cubes.  Using SINFONI observations of the standard stars Hip047903 (G1V) and Hip047235 (G3V), and template spectra \citep{pick98} for their specific spectral types scaled to the appropriate magnitudes in the $H$-band, we determine the relative flux response for each wavelength slice across the datacube, and apply the derived correction (accurate to $\sim 12\%$ in absolute flux values) to the cube data.

\section{Analysis and results}

\subsection{The spatially resolved narrow H$\alpha$ emission}
In order to reveal the underlying kinematics of the host galaxy, we need to extract the extended narrow line emission from our datacube.  We firstly remove the continuum emission in each spaxel via a least-squares fit to a second-order polynomial (excluding the wavelength region covered by the broad H$\alpha$ emission), and then turn to the more difficult task of separating the line emission of the host galaxy from that of the AGN itself.  
As the broad lines originate only from the (unresolved) immediate vicinity of the AGN, the observed spatial variation of broad line intensity in our datacube can be used to accurately constrain the point spread function for our data \citep[as demonstrated by][]{jahn04}. For each spaxel in our data cube, we isolate the wavelength ranges of the spectrum which contain strong broad line emission, but without contamination by sky line residuals, to either side of the spatially extended but spectrally compact narrow line emission (Fig.~\ref{fig1}a).  Using the emission either side of broad H$\alpha$, we subtract any residual background continuum level, and sum up the remaining flux in the selected regions of the broad line. This provides us with a relative measure of the broad line flux on a spaxel-by-spaxel basis in the form of a PSF image which accurately traces the relative distribution of light from the quasar point source at all wavelengths within our datacube (Fig.~\ref{fig1}b). The quasar spectrum itself is well described by the spectrum of the peak spaxel in our datacube, which is dominated by AGN emission alone. We construct a quasar-emission datacube by scaling the quasar spectrum by the relative flux in each pixel of our PSF image.  Subtraction of this datacube from our calibrated data cleanly and accurately removes both the broad and narrow H$\alpha$ emission due to the AGN, and results in a new datacube which contains only host galaxy emission. Non-negativity within the wavelength region covered by the broad  H$\alpha$ emission is not enforced. For twenty spaxels close to the center of the PSF, errors in the subtraction  of the highly luminous QSO emission dominate the resulting spectra and in a few cases result in mild over-subtraction of the QSO emission.  We do not retrieve reliable information on the narrow H$\alpha$ emission from these spaxels.

\begin{figure}
\vspace{-2cm}\includegraphics[width=\columnwidth]{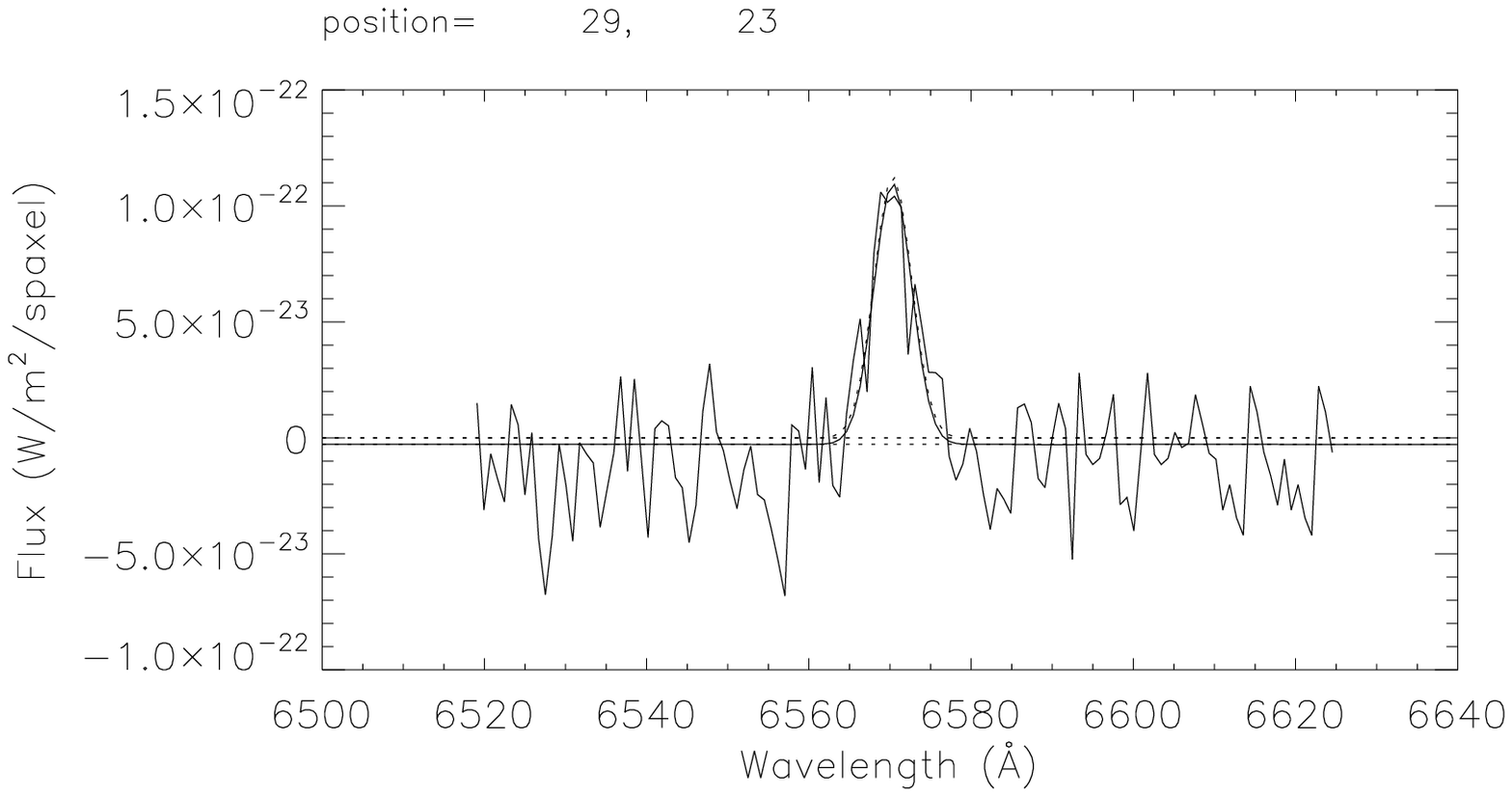}\vspace{-2cm}\\
\includegraphics[width=\columnwidth]{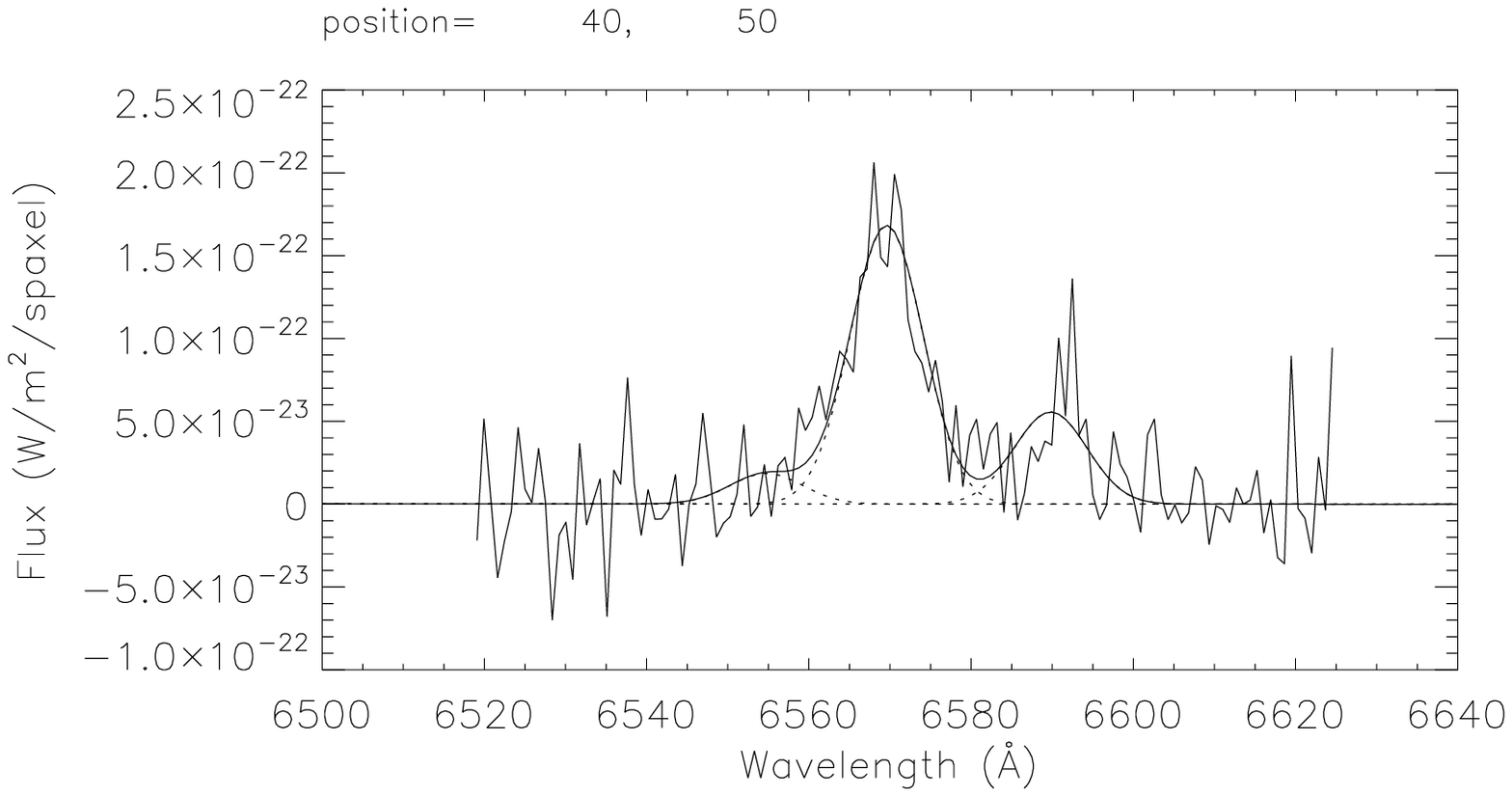}\vspace{-2cm}\\
\includegraphics[width=\columnwidth]{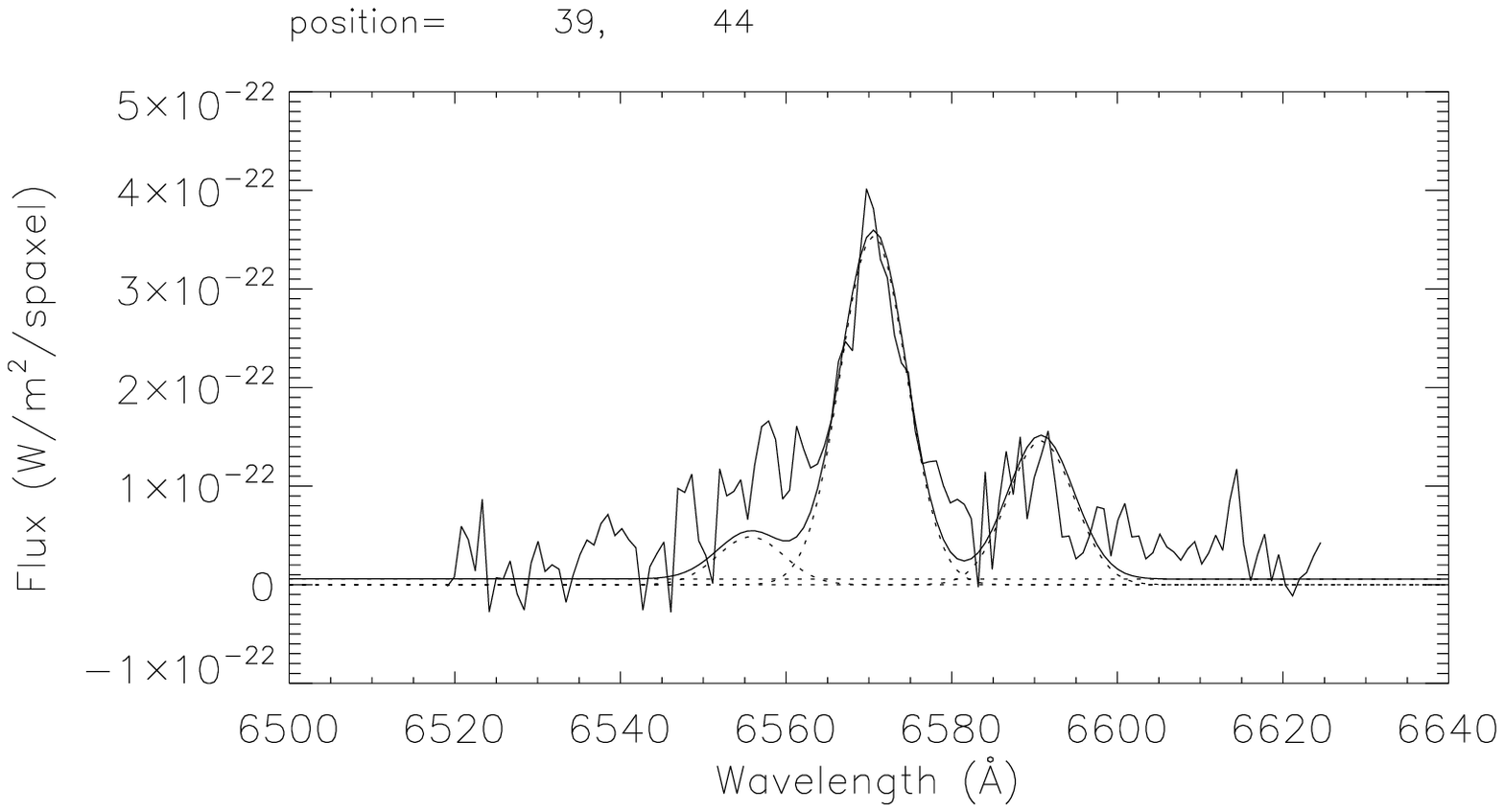}
\caption{Example fits to the narrow line emission in the post-QSO subtraction datacube for three different spaxels, one in the tidal feature to the south east (top), one to the north (center), and one in the central region (bottom). The labelled positions are in the same coordinates as shown on Fig.~1b and all subsequent 2-d images of the field of view. Gaussian line profiles plus a linear continuum are used to simultaneously fit the narrow H$\alpha$ and [N\textsc{ii}]6548,6584 emission lines in the source rest-frame, assuming the canonical SDSS DR3 redshift of $z=1.31177$. The weaker [N\textsc{ii}] lines cannot be accurately identified in all spaxels, and in such cases are excluded from the fit.
\label{fig2}}
\end{figure}

\begin{figure*}
\includegraphics[scale=0.5]{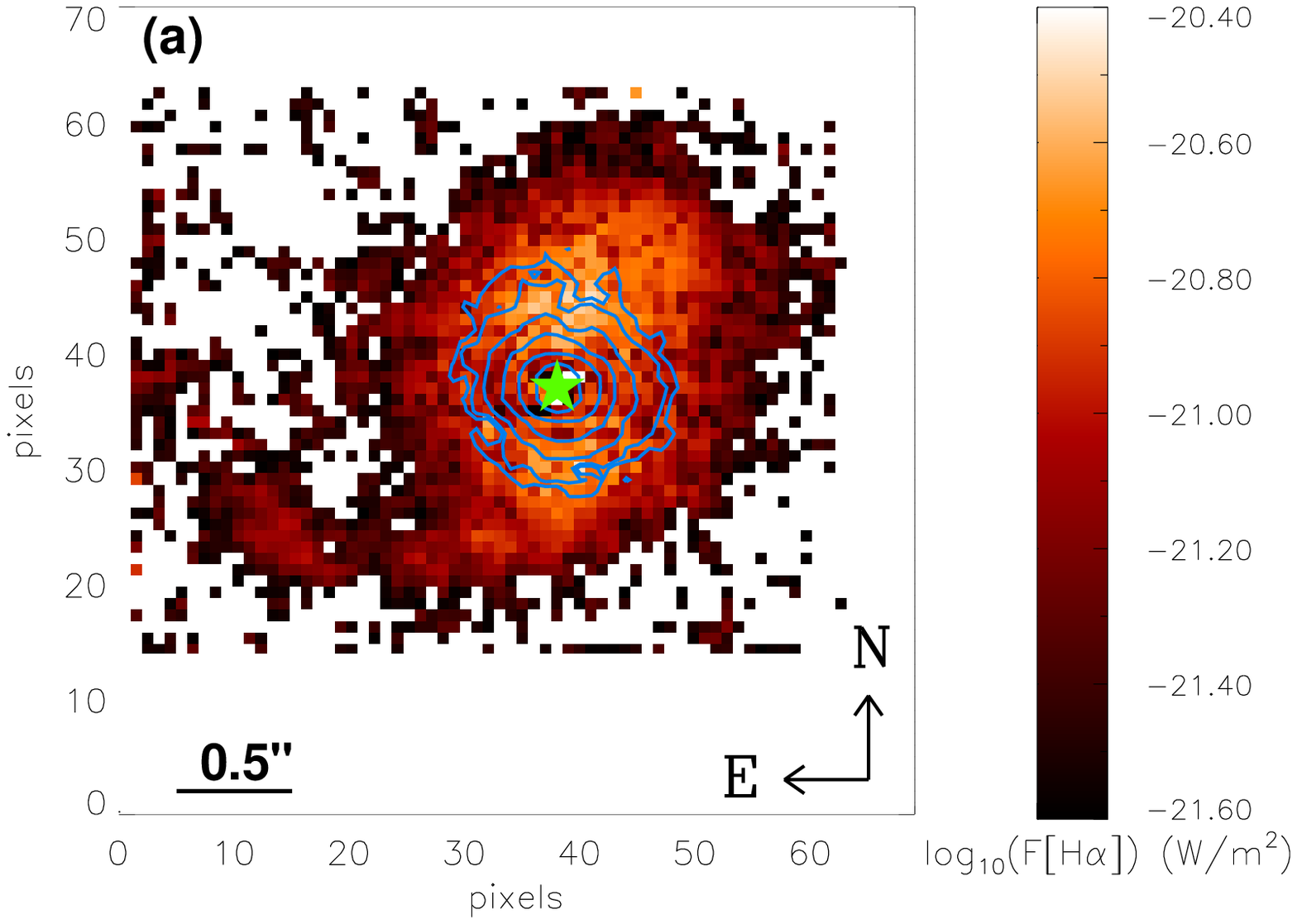}
\includegraphics[scale=0.5]{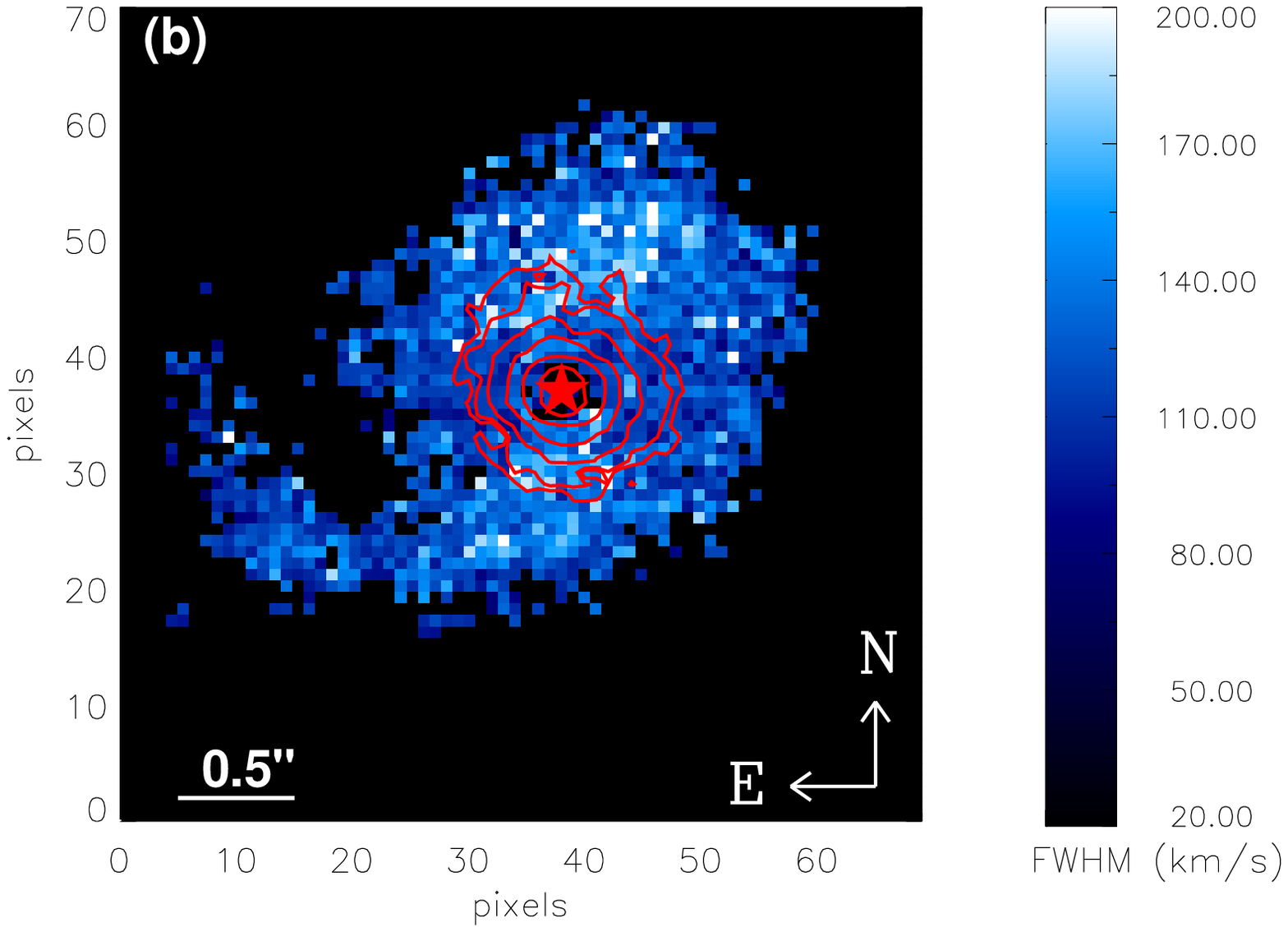}
\includegraphics[scale=0.5]{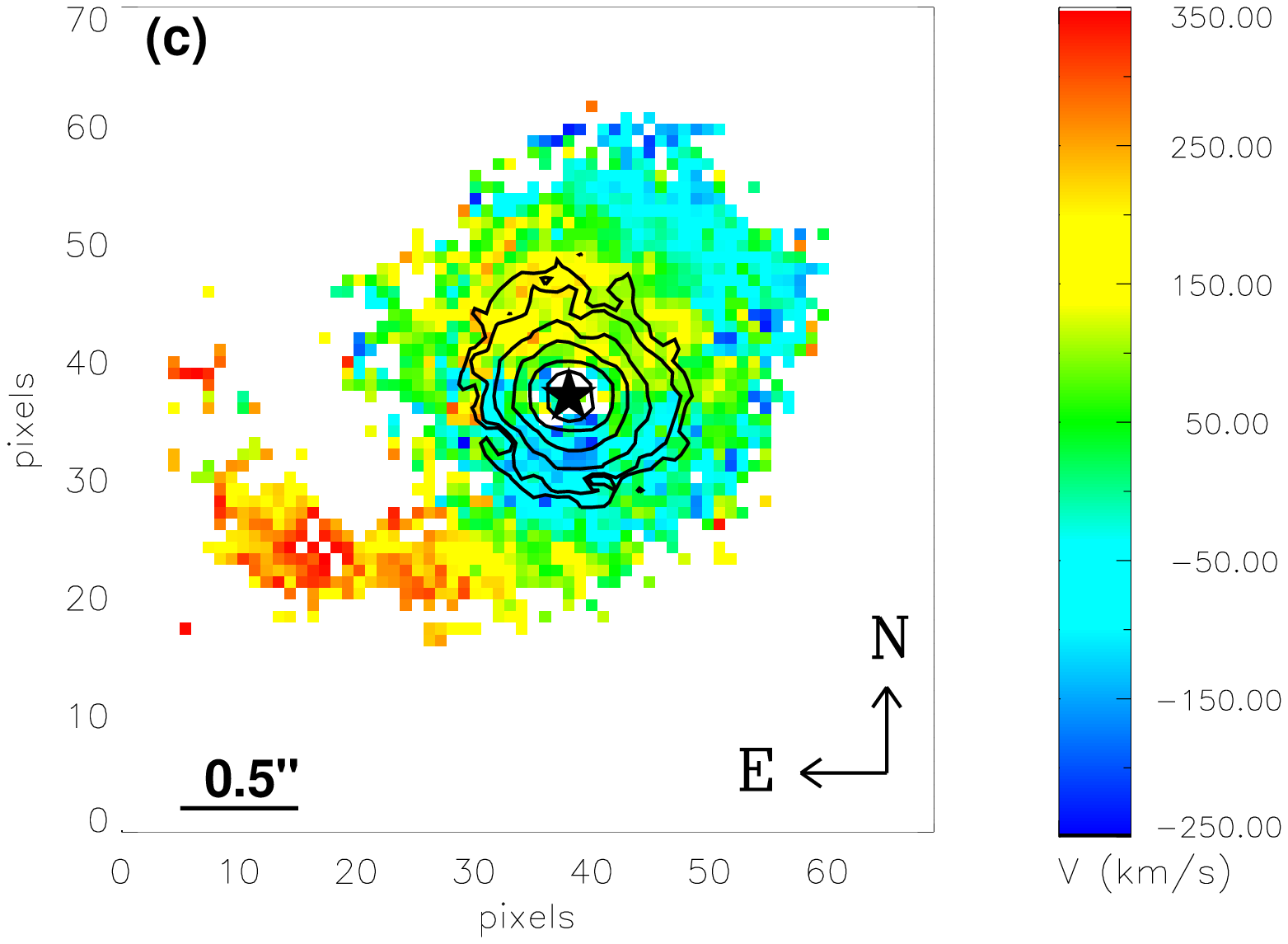} 
\includegraphics[scale=0.5]{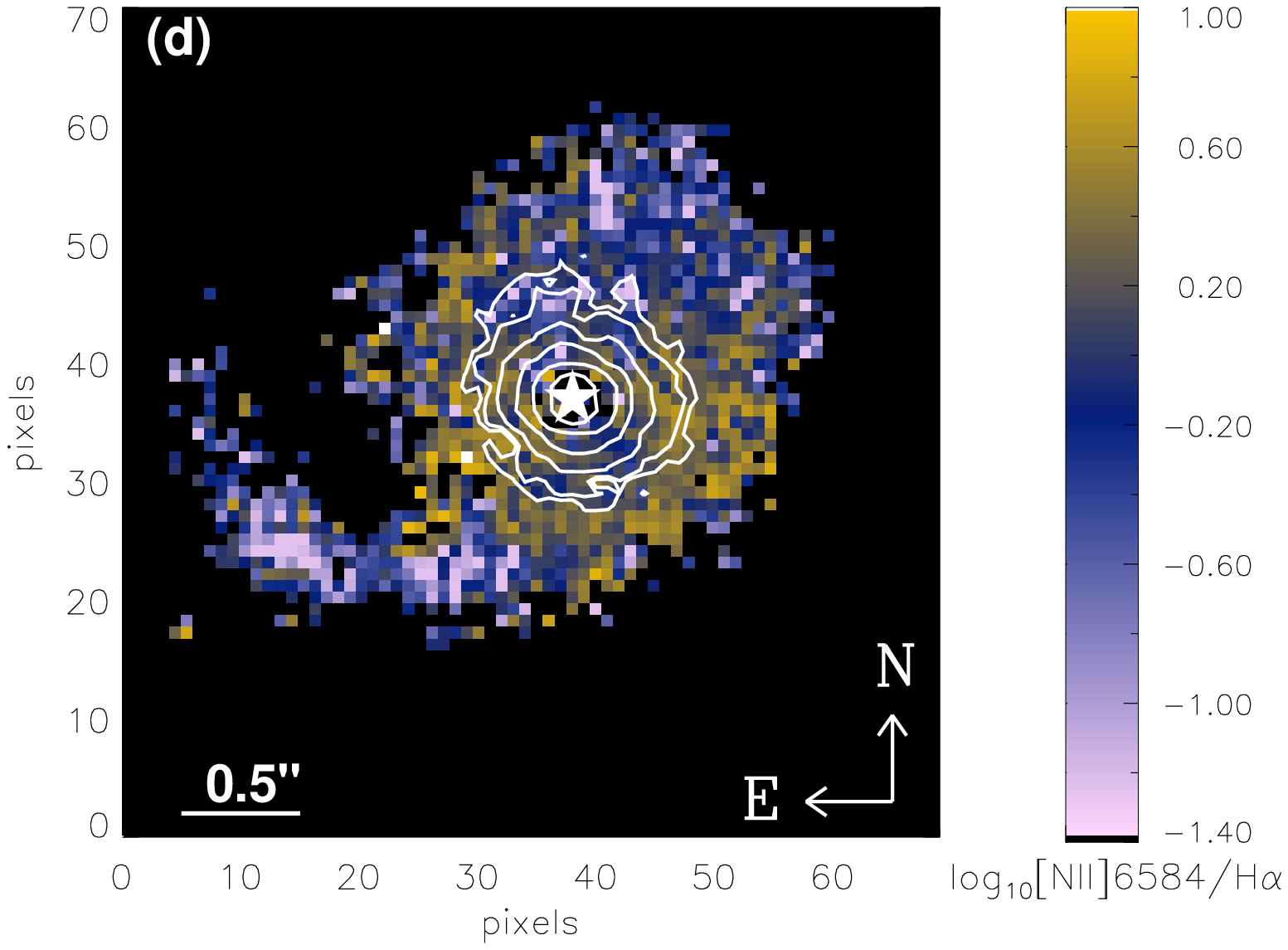}
\caption{Results of emission line modelling (see section 3.1), overlaid with PSF contours from Fig.~1b. (a - top left) Narrow $H\alpha$ emission line flux image. (b - top right) Line width of modelled narrow line emission. (c - bottom left) Narrow line emission velocity field. Note the change in the rotation direction between the inner and outer regions. Zero velocity is defined by the center of rotation in the central regions. (d - bottom right) [N\textsc{ii}]6584/H$\alpha$ emission line ratio. 
\label{fig3}}
\end{figure*}

\begin{figure*}
\includegraphics[scale=0.21]{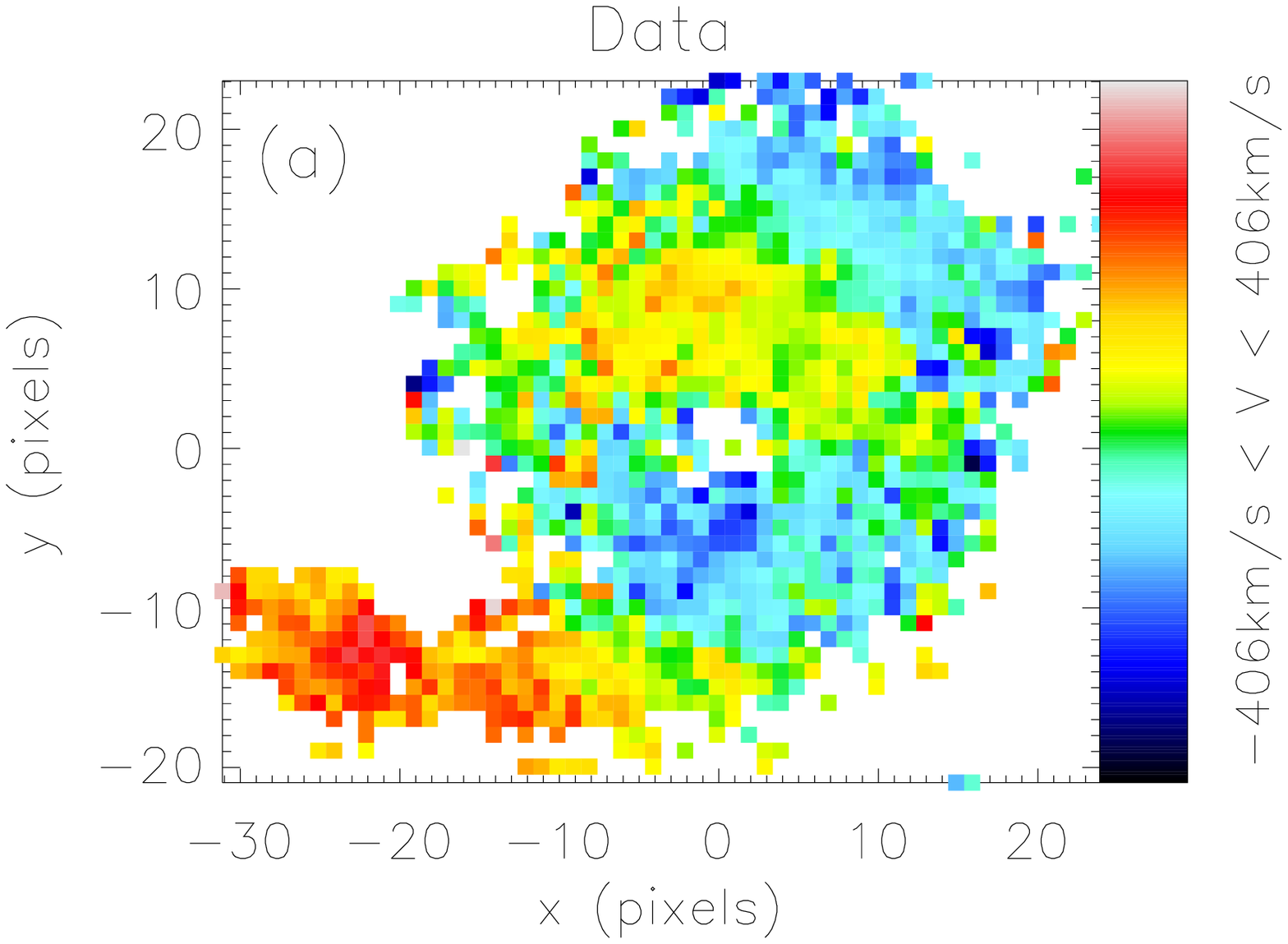}\hfill
\includegraphics[scale=0.21]{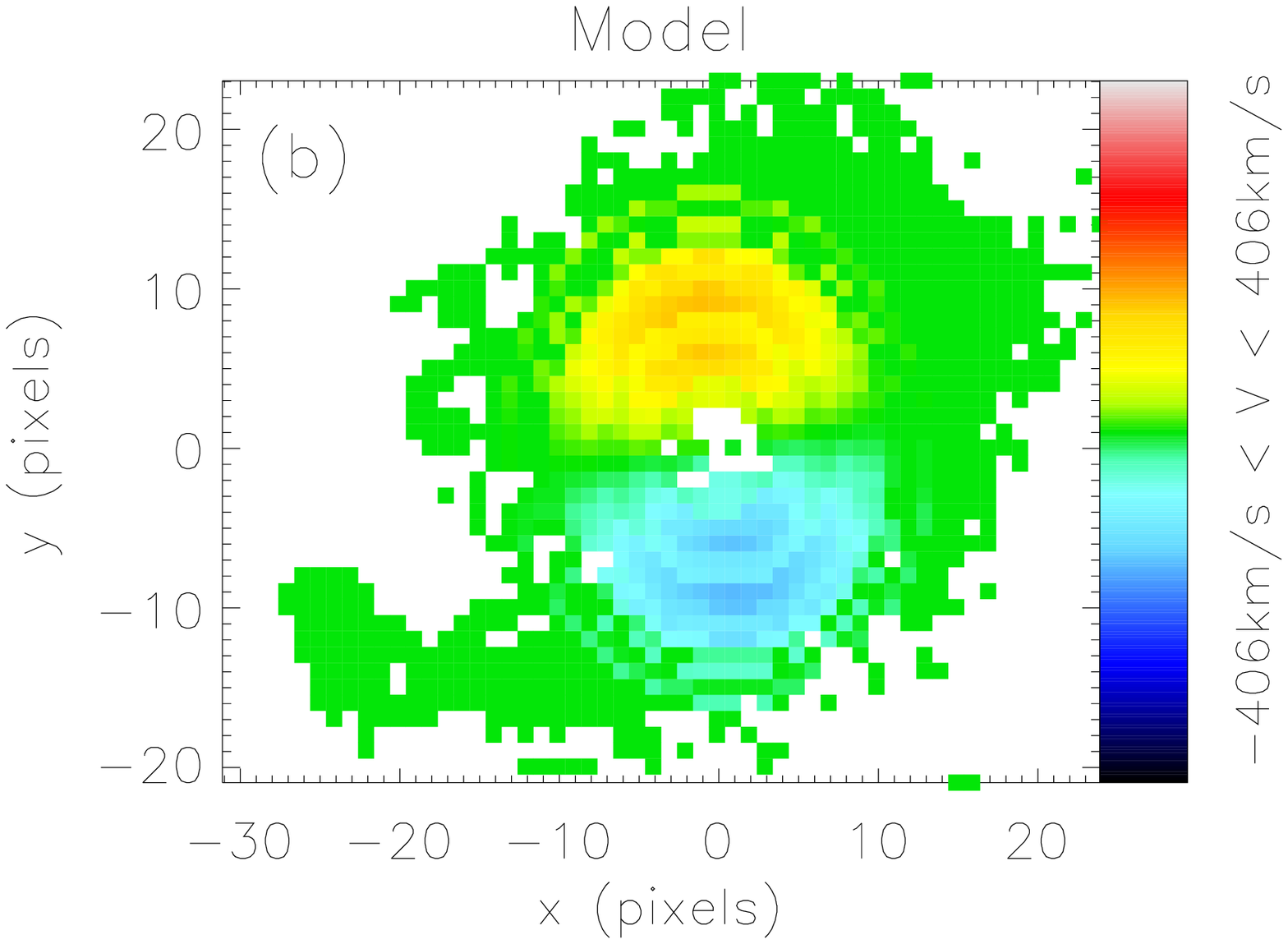}\hfill
\includegraphics[scale=0.21]{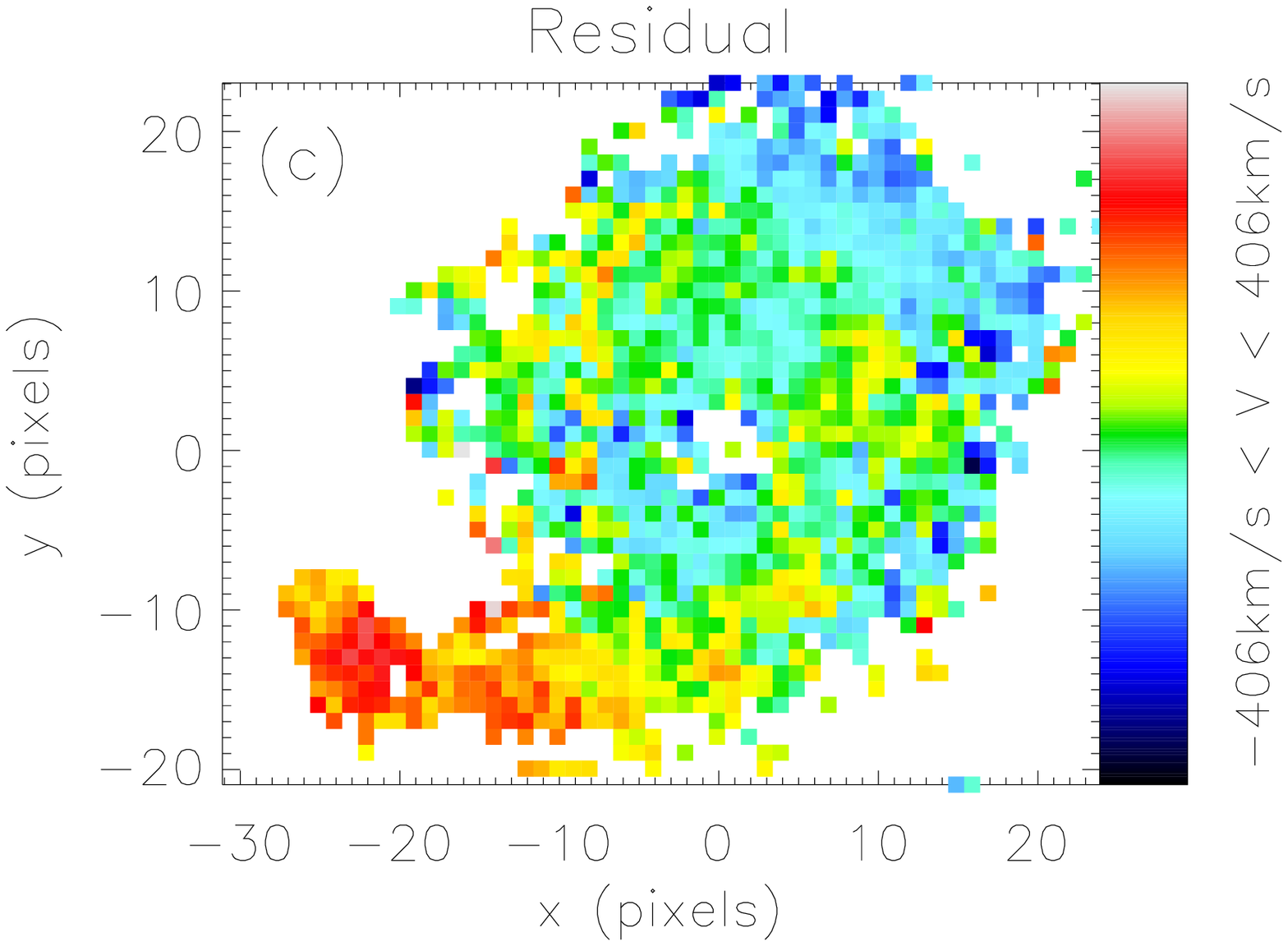}\hfill
\includegraphics[scale=0.21]{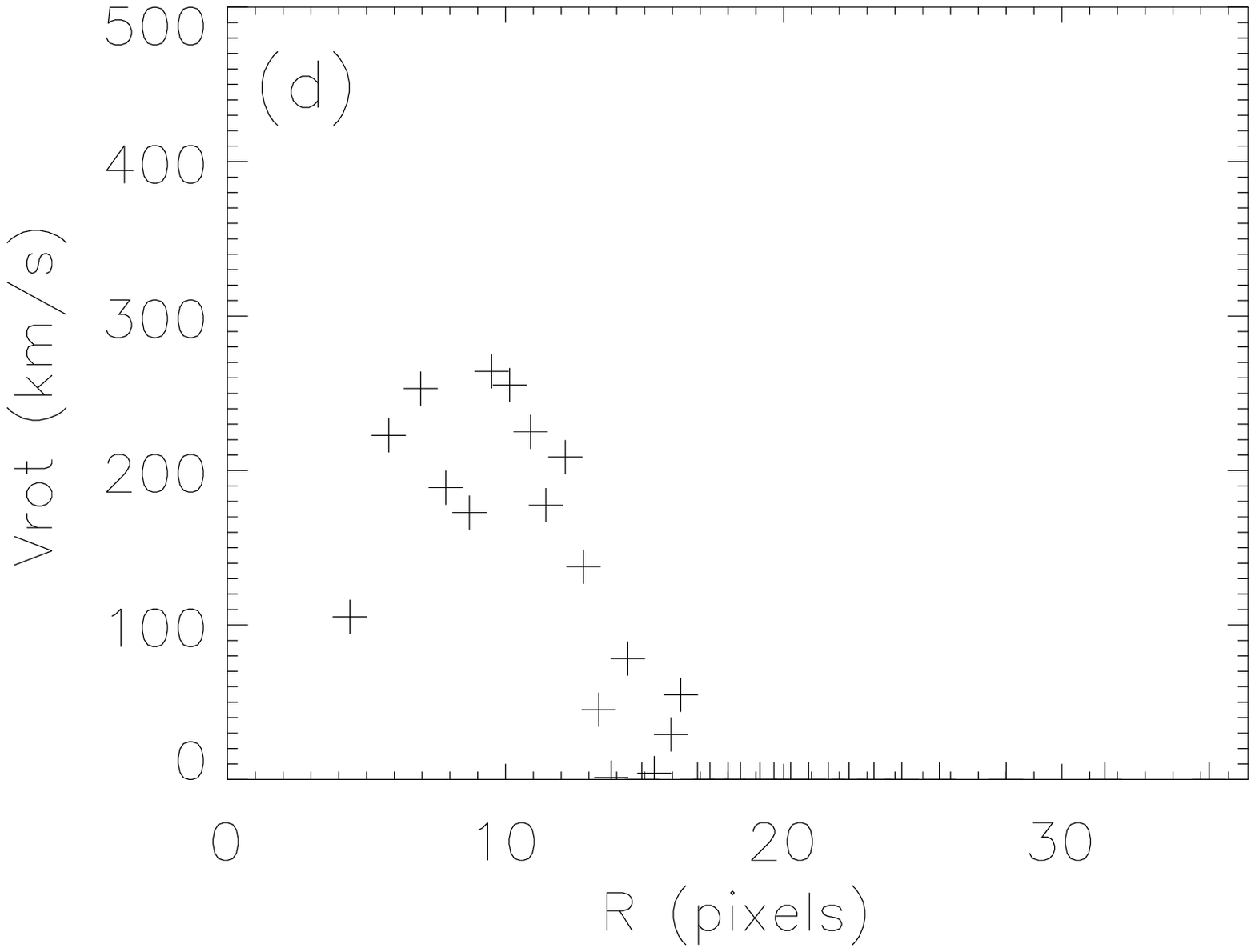}\\
\includegraphics[scale=0.21]{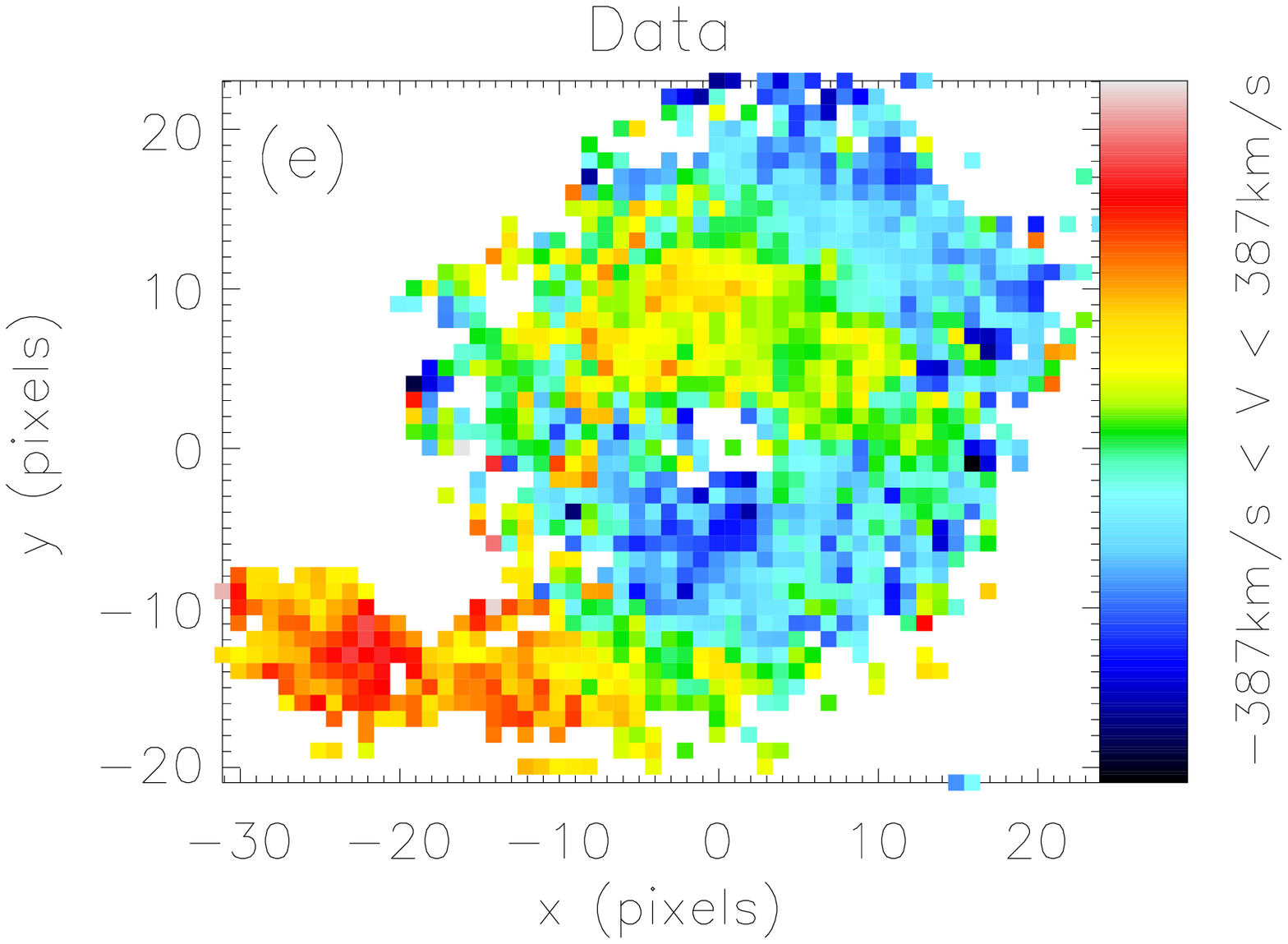}\hfill
\includegraphics[scale=0.21]{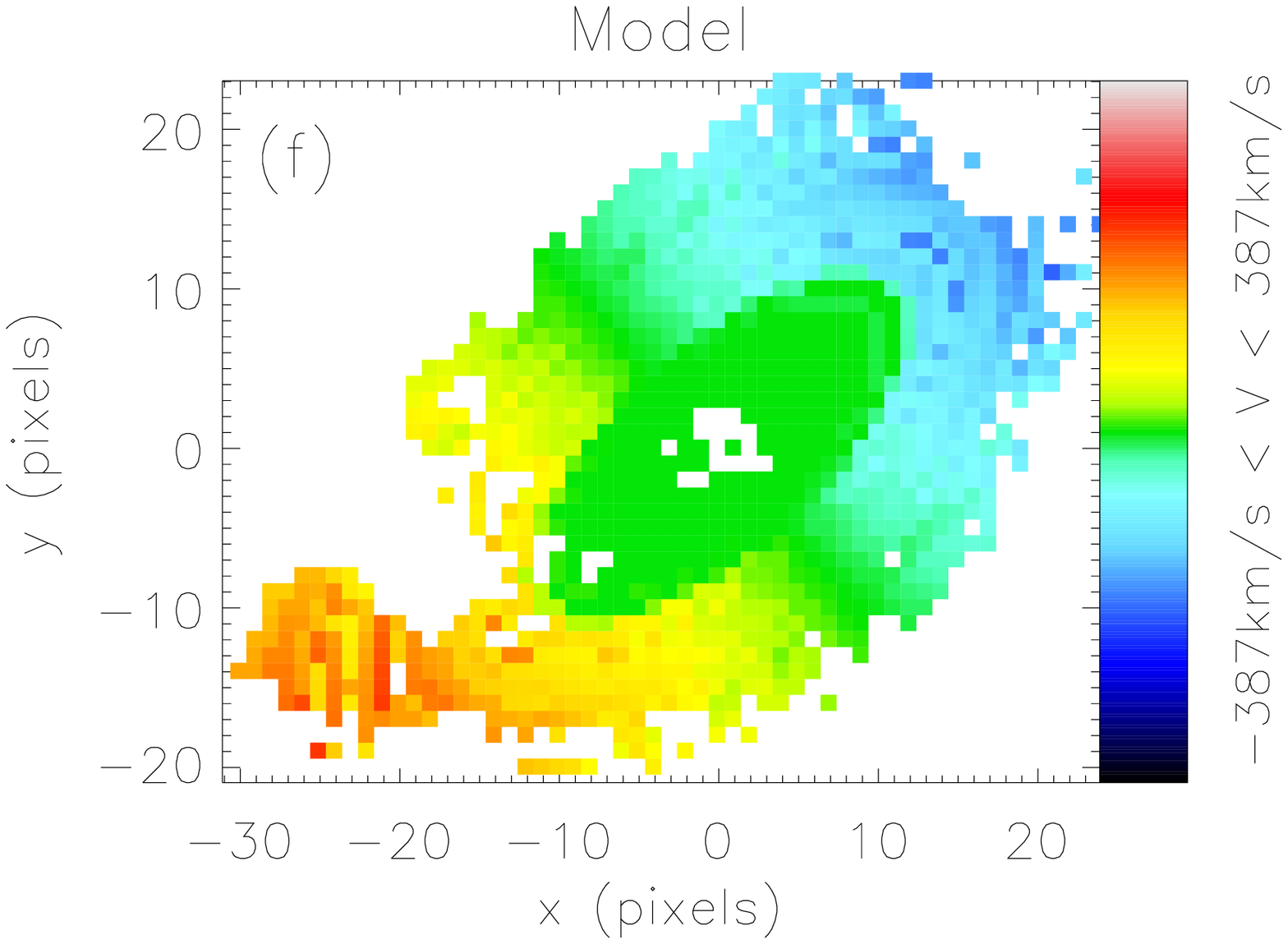}\hfill
\includegraphics[scale=0.21]{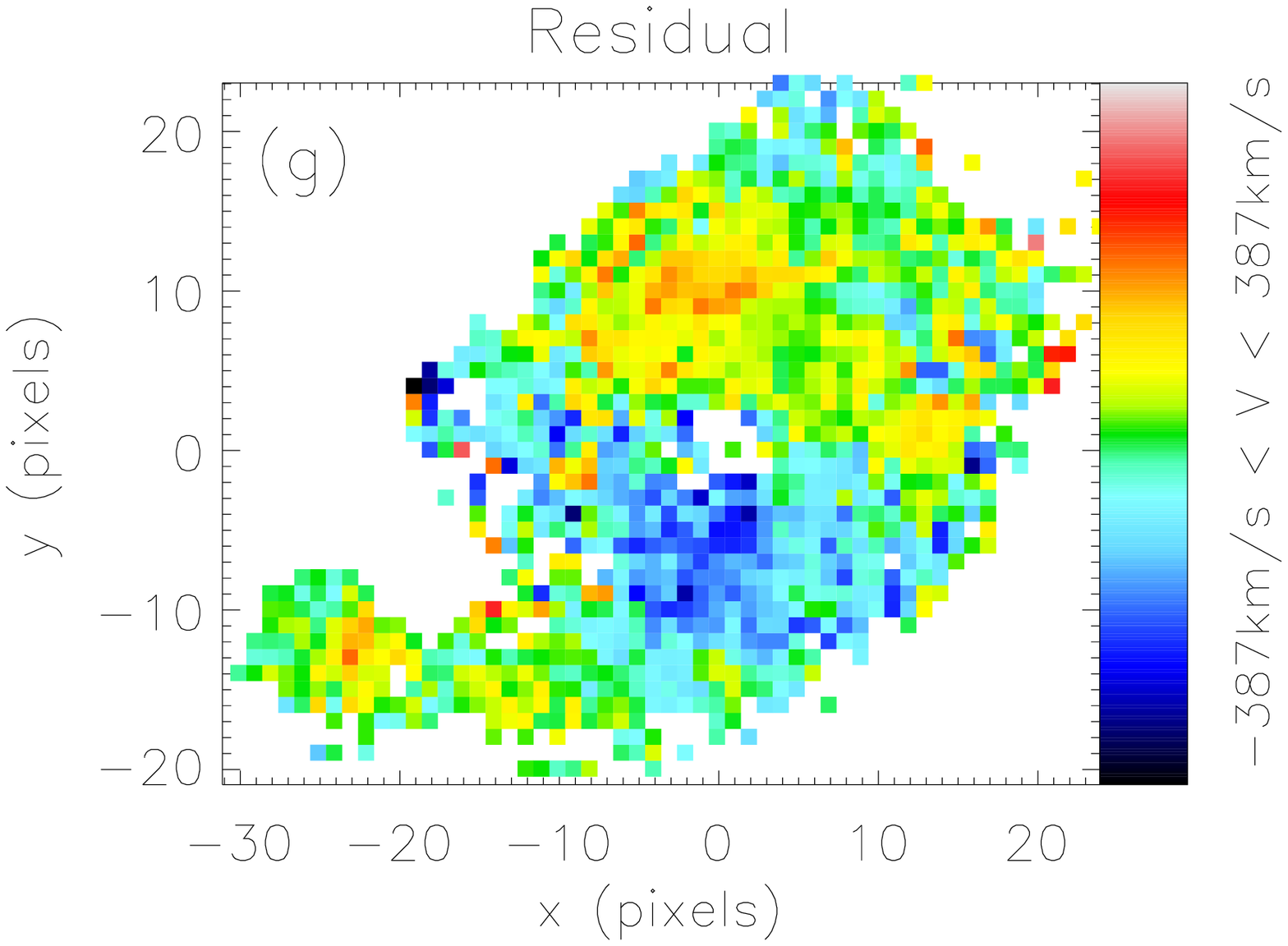}\hfill
\includegraphics[scale=0.21]{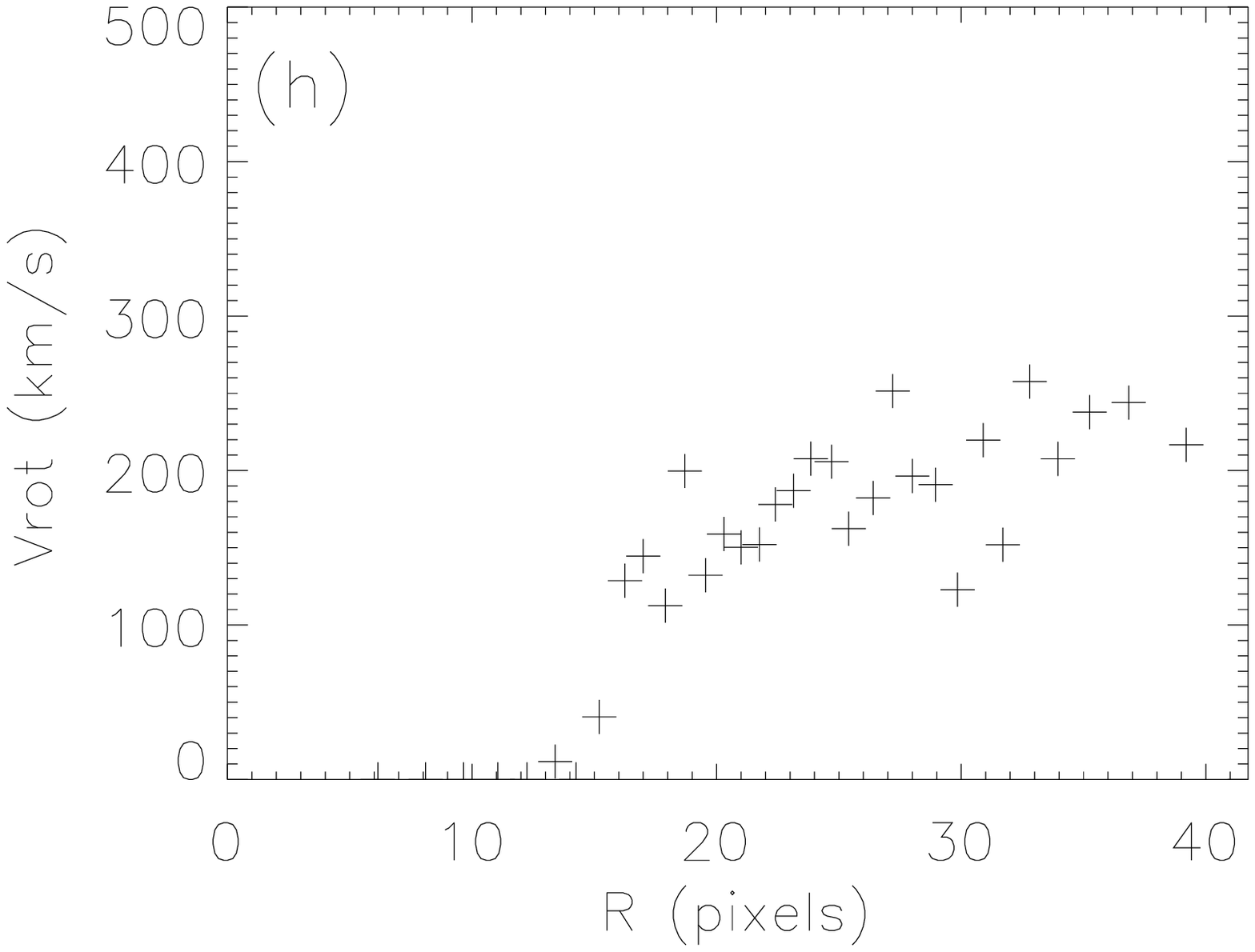}
\caption{Projected velocity field and empirical first-order modelling. Here we examine separately the two major velocity structures in the emission line gas, prior to more extensive multiple-component modelling (Figs.~5-7). Harmonic decomposition techniques are used to fit circular rotation along two sets of ellipses of fixed position and inclination angles (see section 3.2.1). We display the data (frame a), the overall model velocity field for the first set of simple tilted ring models for which the  position angle was tuned to target the central rotating region (frame b), model-subtracted residuals (frame c) and the resulting velocity as a function of radius for each extracted elliptical annulus, for this first set of fits (frame d). Rotational velocity increases to a maximum of $220\pm50 \rm km \, s^{-1}$; at larger radii (beyond a radius of $\sim 12$ pixels) it then declines steadily due to the influence of the counter-motions in the tidal features.  Panels in the bottom row display the data (frame e), the model velocity field for the second set of fits with a position angle tuned to target the tidal feature (frame f), residuals (frame g) and the model velocity profile as a function of radius (frame h).  In the region dominated by the tidal features, velocity varies roughly linearly with radius.
\label{fig4}}
\end{figure*}

 For each spaxel of the QSO-subtracted datacube, we then fit the data with a linear continuum and a simultaneous combination of three gaussians of the same width and fixed relative centroid but varying strength for the H$\alpha$ and [N\textsc{ii}]6548\AA,6584\AA\ emission lines. The spectra are converted to the source rest frame using the canonical SDSS DR3 redshift of $z=1.31177$. Fig.~\ref{fig2} displays example fits to three different spaxel spectra in the QSO-subtracted datacube from different regions of the field of view. 
The results are then masked to exclude unreliable fits to noise rather than to genuine line emission, i.e. excluding the outermost rows and columns of data where no genuine emission features are observed, spaxels where the fit results in extreme velocities ($-100 < v < 600 \rm km\ s^{-1}$ without subtraction of any further systemic velocity at the assumed redshift of 1.31177, confirmed as fits to noise features via visual inspection), line centroid uncertainties $>40 \rm km\ s^{-1}$, and H$\alpha$ fluxes of $<10$ counts or uncertainties of $>3$ counts per binned spaxel.  Fig.~\ref{fig3} displays the extracted H$\alpha$ flux image (Fig.~\ref{fig3}a), line width image (Fig.~\ref{fig3}b), velocity field after correcting the systemic velocity (Fig.~\ref{fig3}c)  and [N\textsc{ii}]6584\AA/H$\alpha$ line ratio map (Fig.~\ref{fig3}d), overlaid with the PSF contours from Fig.~\ref{fig1}b.  A more in-depth analysis of the emission line properties of this system is deferred to our study of a larger sample of similar objects.
In our subsequent modelling of the velocity field, we further mask the outermost, low signal-to-noise regions, via boxcar-averaging the line flux in 9 by 9 pixel regions and excluding regions of low average signal. The remaining unmasked pixels are illustrated in the line flux image displayed in Fig.~\ref{fig4}a.

\subsection{Velocity field analysis}
As is clear from Fig.~\ref{fig3}c, this source displays a complex velocity field. A quick visual inspection suggests the central region can be well described by ordered disk-like rotation of the gas, with a reversal of the gas velocity at larger radii along at least one arm-like feature, and possibly a second outside our field of view to the north. With a pixel scale of $\sim 0.4\,$kpc per pixel, the southern arm extends to a projected distance of approximately 20\,kpc, suggesting a tidal origin for this feature as the most likely scenario.  Although there are no radio observations made specifically of this object, the fact that it is not a FIRST \citep{first1, first2} radio source suggests that no strong radio jet is present in this source, and thus that jet-driven shocks/interactions are unlikely to be a significant factor for the observed kinematics of this object 

Our aim is to use this observed gas velocity field as a tracer of  the overall gravitational potential of the galaxy. The fact
that disk-like rotation is present in the gas velocity field, as we will show in
more detail in the following sections, is independent of the geometrical structure 
of the underlying stellar body and makes no statements about the stars
being in either a disk-dominated or bulge-dominated structure\footnote{We note that in spite of the depth of our observations, there is insufficient extended continuum flux within the datacube to extract any information about the host galaxy morphology.}.  Gas disks can be commonly observed in elliptical galaxies \citep{ost10, ost07, emonts08}, and it is often the case that their kinematics do not trace those of the underlying stellar population. Indeed, the velocity structures may be highly decoupled, both in the case of active \citep[e.g. HE1029-1408 at $z=0.086$, which displays low stellar rotation but substantial gas rotation; ][]{huse10} or inactive \citep[e.g. NGC2768, NGC4314, NGC 4526 and NGC 4546; ][]{sarzi06} galaxies.

In order to properly characterise the  rotation of the central regions and to extract a dynamical mass, it is important that we can also simultaneously account for the contrasting peculiar velocity structures dominating the gas velocity field at larger radii ($> 5$kpc).
However, the signal-to-noise of the line emission in the individual spaxels where these two velocity structures overlap is not sufficiently high to separately resolve more than one line profile, nor can multi-spaxel extractions allow the fitting of multiple line profiles without unavoidable degeneracies in the relative line widths, centroids and strengths. We therefore choose to use multi-component fits to the full velocity field.

\subsubsection{Separation of emission from the central and outer tidal components}
In the process of producing a model velocity field, it is essential to consider the strength of the contribution of line flux at different velocities from adjacent pixels during convolution with the observational PSF. The observed velocity in a given position is not simply the convolution of the model velocity field with the observational PSF centered on that location: as the line flux varies across the observational field of view, the proportional contribution of different pixels to each specific value in the output model velocity field will also vary. In order to correctly weight each velocity component of our model during PSF convolution, we also need to include a model of the relative pixel-to-pixel line flux -- for both the central rotating disk and also the additional tidal arm features. 

In producing a model for the velocity field of J090543.56+043347.3, we start by examining the two major velocity structures separately, followed by more extensive multiple-component modelling. We obtain additional constraints on these velocity structures via the application of harmonic decomposition \citep[cf.][]{vdven2010, fathi05}. 
 This treatment divides a velocity field and associated signal-to-noise map into elliptical annuli of differing semi-major axes but identical orientations, inclinations, ellipticities and centroids, and iteratively extracts the velocity profile as it appears within each annulus as a finite number of harmonic terms. Velocity and higher-order moments can then be readily characterised, and the nature of any deviations from non-circular motion assessed.  The same tools can also be used to model a velocity field within a succession of elliptical annuli, while also varying the parameters listed above for each annulus.

 We use this technique to confirm the centroid of our velocity field, and find that the central regions are indeed well matched by ordered rotation, oriented at an angle of $\sim 10^{\circ}$ east of north, and inclined to the line of sight at an angle of $35 \pm 15^{\circ}$. The results of fitting the velocity field with a single simple inclined rotating disk, of fixed position angle and inclination angle, and a velocity which varies with radius, are displayed in Figs.~\ref{fig4}a-d. These plots illustrate the velocity field, model, residuals, and radial rotation profile of the model respectively (note that each data point in Fig.~\ref{fig4}d is derived from all spaxels which contribute to the model elliptical annulus at that radius). Velocity increases rapidly in the central regions to a maximum of $220\pm50 \rm km\,s^{-1}$, and then declines rapidly beyond a radius of $\sim 12$ pixels due to the influence of the counter-motions in the tidal features (Fig.~\ref{fig4}d).

We also carry out similar simplistic modelling of the velocity field targeting the outer regions. The velocity field due to the tidal arm features is oriented at an angle of approximately 140$^{\circ}$ offset from the central structures, varies relatively monotonically with radius, and dominates the velocity field beyond a projected radius of approximately 5kpc (= 12 pixels; Fig.~\ref{fig4}e--\ref{fig4}g). Full harmonic decomposition along each set of the defined ellipses implies that neither the central rotating structure nor the outer tidal arms show any strong evidence for more complex motions within the regions they respectively dominate; the inclusion of any further velocity structures in our multiple-component modelling is therefore not required. 

\begin{figure*}
\includegraphics[scale=0.272]{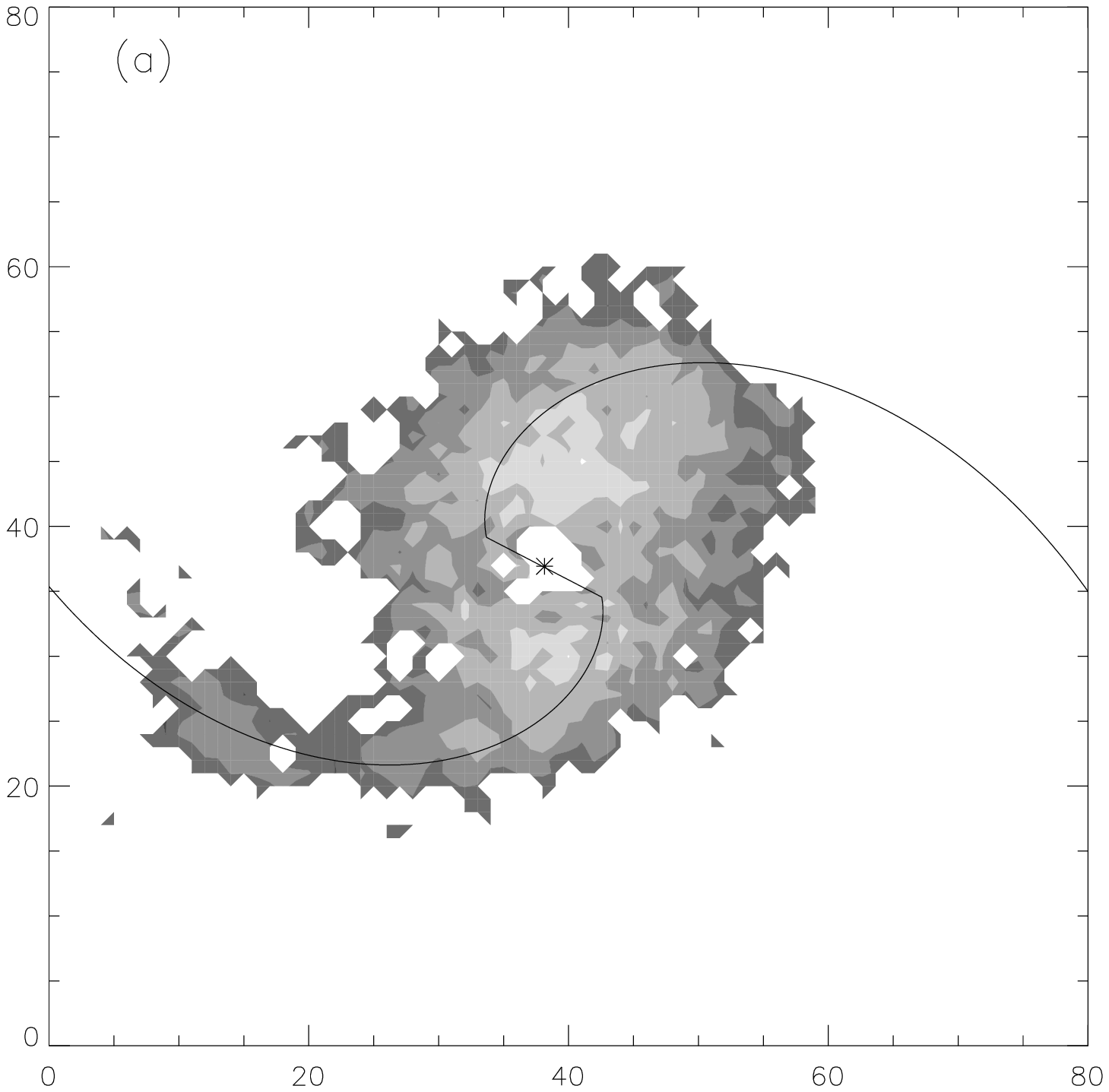}\hfill
\includegraphics[scale=0.272]{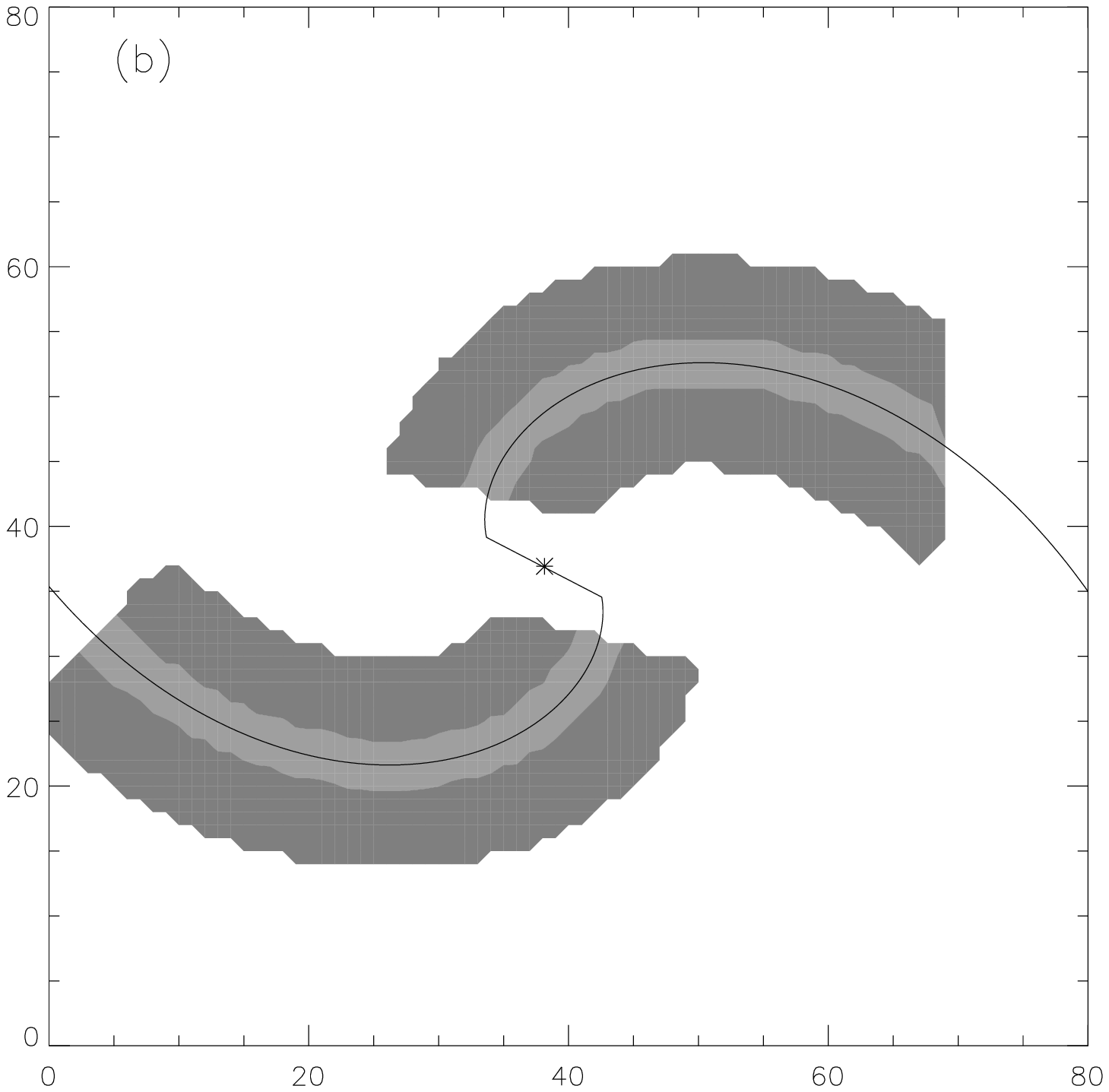}\hfill
\includegraphics[scale=0.272]{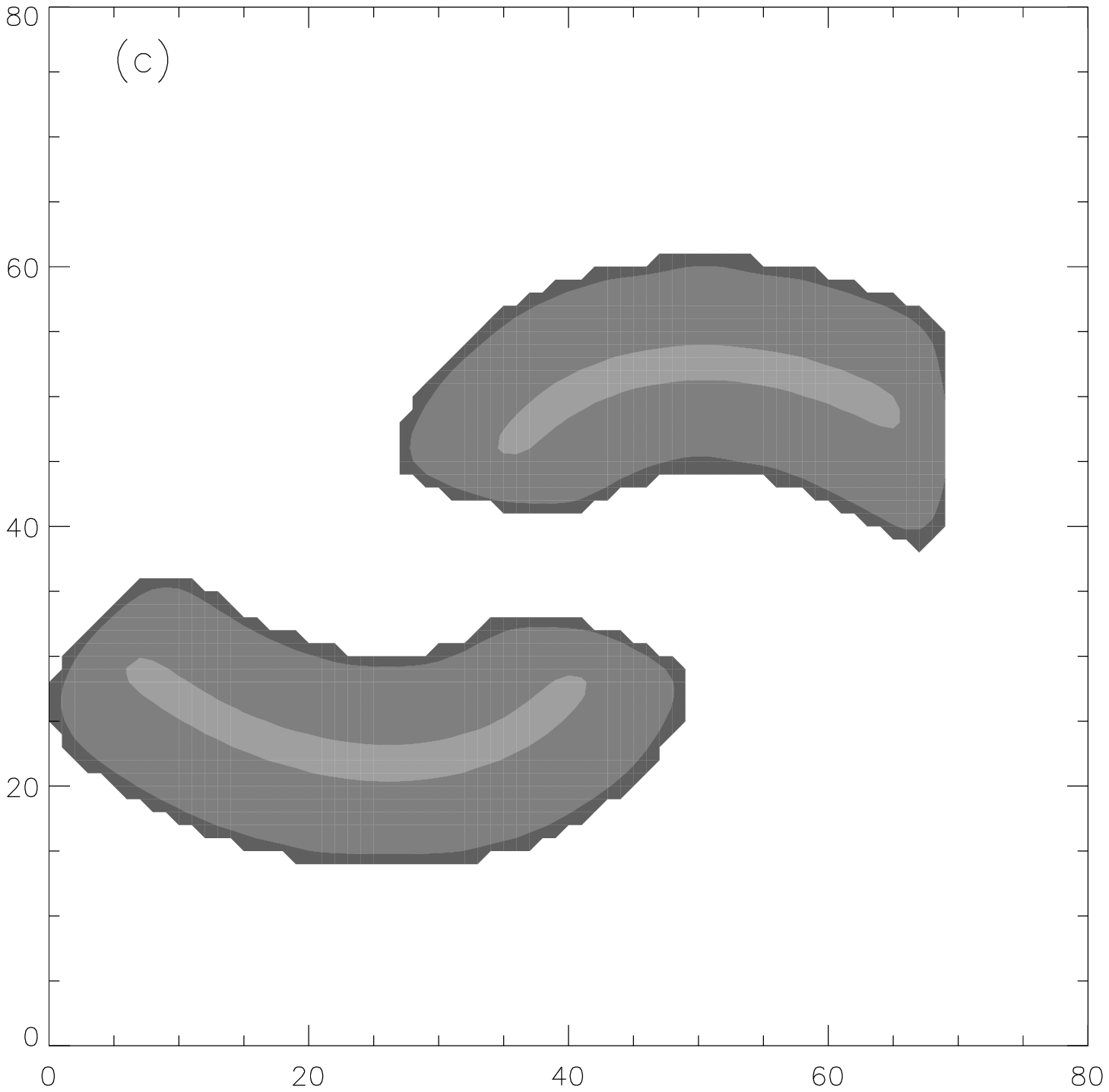}\\
\includegraphics[scale=0.272]{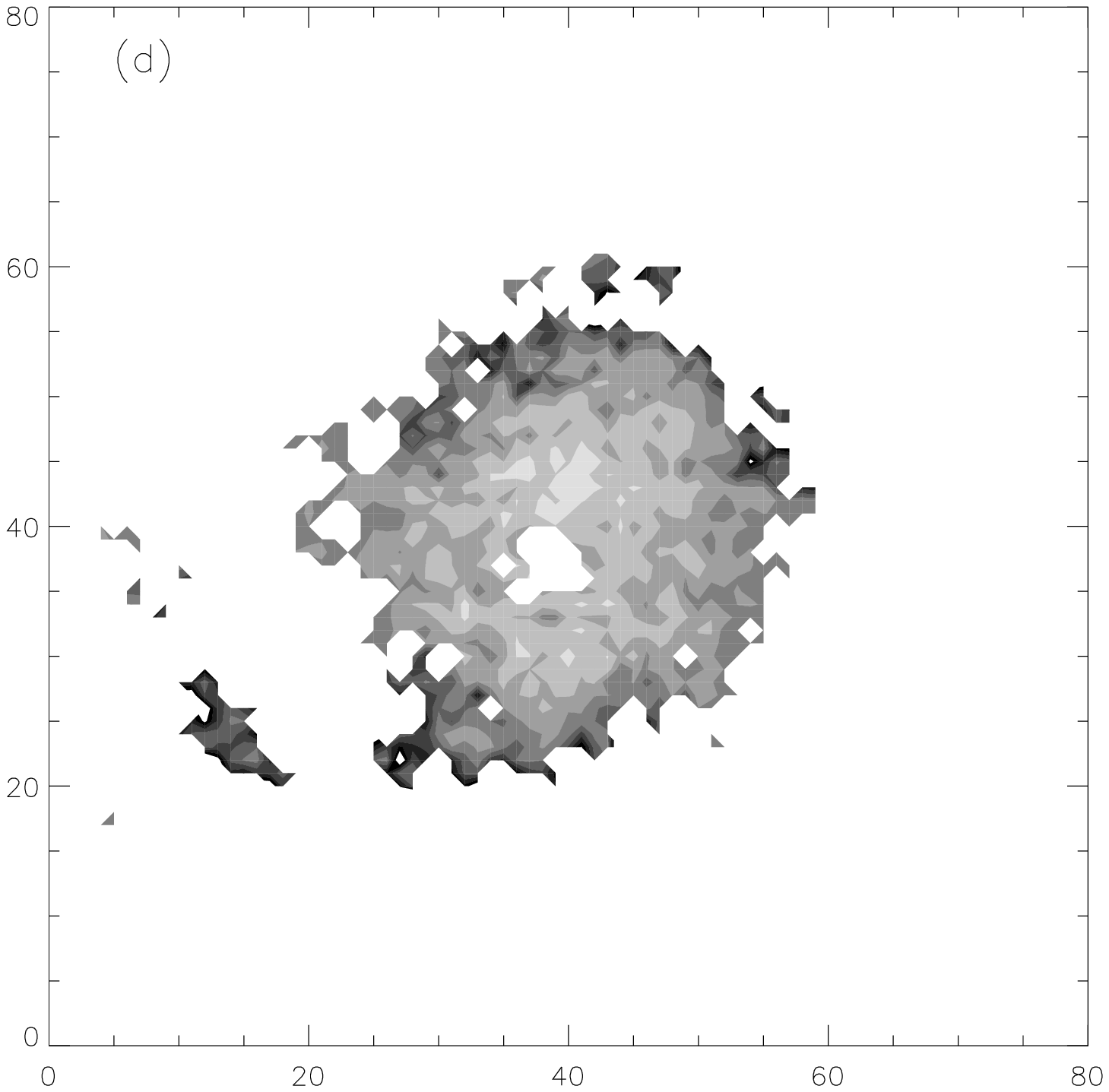}\hfill
\includegraphics[scale=0.272]{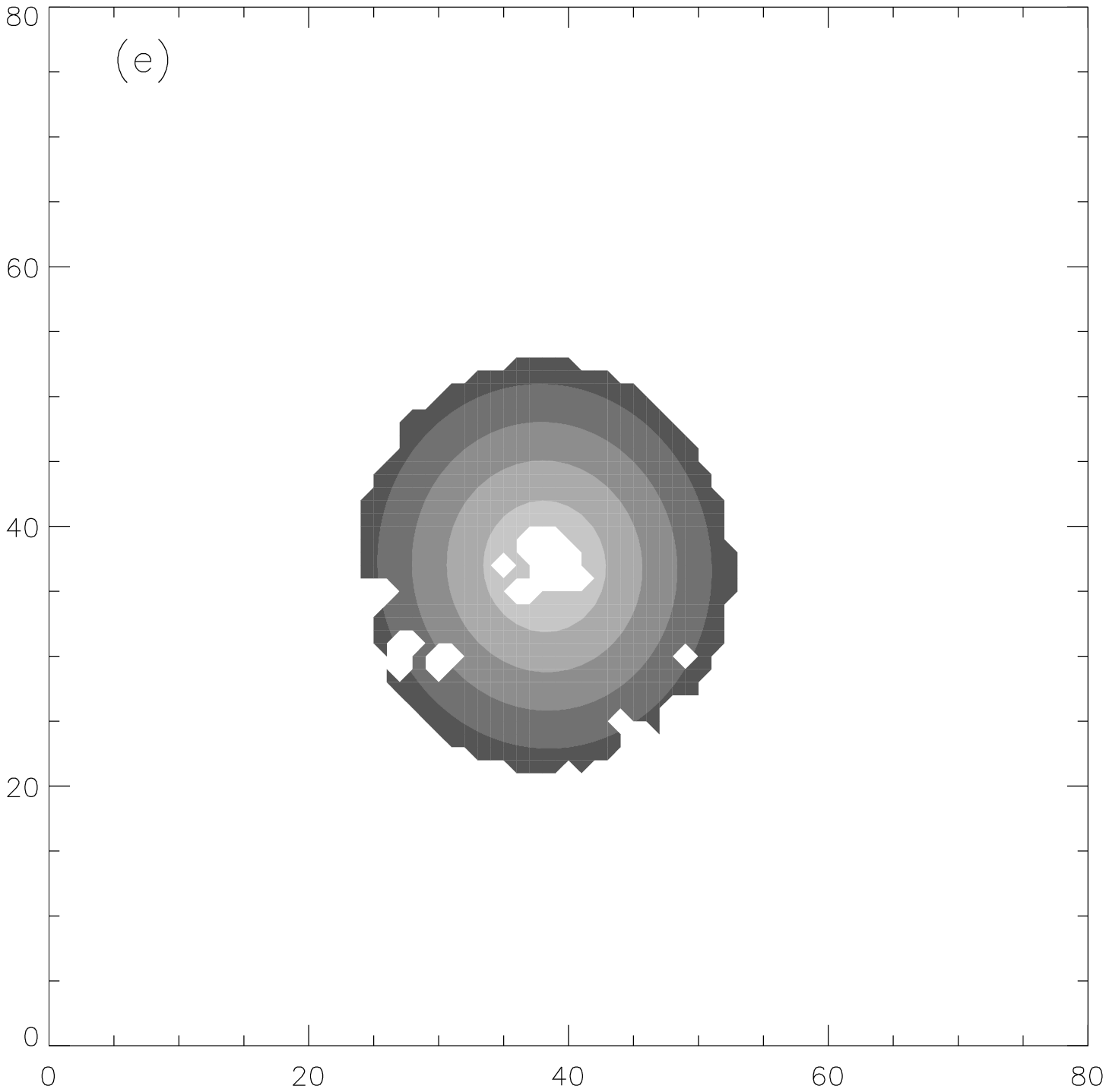}\hfill
\includegraphics[scale=0.272]{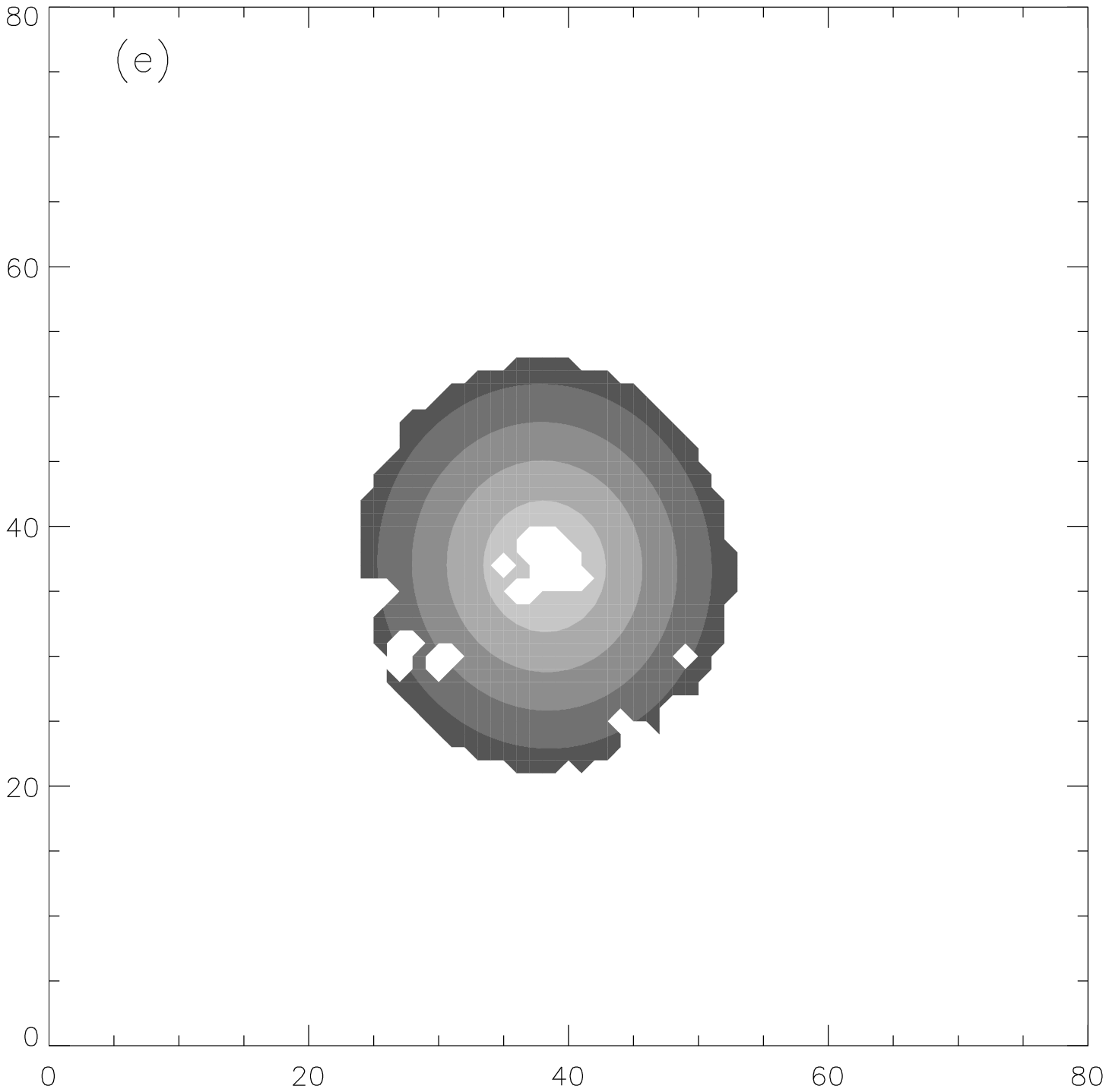}
\caption{Modelling the $H\alpha$ flux distribution (see section 3.2.1) as a prerequisite to full flux-weighted and PSF-convolved multiple-component velocity modelling (results of which are displayed in Figs.~6-7). (a - top left) Narrow $H\alpha$ emission line flux image, plus logarthmic-spiral fit tracing the tidal arms. (b - top center) Model cross-sectional flux profile for tidal arms. (c - top right) Convolution of cross-sectional flux profile with observational PSF, providing best-fit model to narrow line flux in tidal arms. (d - bottom left) Residual narrow H-alpha emission line flux image, after removal of modelled flux in the tidal arm features. (e - bottom center) Convolution of n=1 sersic disk model with the observational PSF, within the orientation and inclination constraints supplied by the velocity field, which provides the best-fit model to the residual line flux. (f - bottom right) Combination of disk and tidal arm flux models, unconvolved with observational PSF, used in flux weighting of velocity models.
\label{fig5}}
\end{figure*}

While harmonic decomposition provides a rapid means of investigating the different types of velocity structure contained with the velocity field of this galaxy, and determining better constraints on the system's inclination angle, we still need to model both structures simultaneously, particularly in order to determine the underlying rotational velocities in the overlap region.
We model the light distribution of the tidal arm features by first tracing the observed line flux with a logarithmic spiral (Fig.~\ref{fig5}a). The average cross-sectional flux profile is determined and applied (Fig.~\ref{fig5}b), and subsequently convolved with the PSF (Fig.~\ref{fig1}b) to create a good match to the observed flux (Fig.~\ref{fig5}c). We then model the residual flux (Fig.~\ref{fig5}d) with an inclined exponential disk (oriented to match the central ordered rotation region of the velocity field), also PSF-convolved (Fig.~\ref{fig5}e), to provide a first approximation to the flux-profile of the rotating disk. (Note that the residual flux after tidal-arm subtraction was better described as an exponential disk than as an $n \ge 2$ S\'ersic profile, as expected for a rotating gas structure.) Fig.~\ref{fig5}f displays our combined relative flux model for the disk and tidal arm features, unconvolved with the observational PSF.  While this does not perfectly match the observed flux profile of the narrow line emission, it does provide a very good approximation to the relative pixel-to-pixel contributions to the velocity field.

\subsubsection{Velocity field model and dynamical mass}

The velocity profile of the arm features is modelled as a linear change in velocity with location along the arms, defined from the outer regions free of any influence from the central rotating disk, and extrapolated inwards from there.
For the central rotating disk, we assume for our model velocity field a Persic universal rotation curve \citep{per96} as a generally valid example of the rotation curve for a disk. This can be described as
 \begin{multline}
V(x) =  V(R_{opt}) \left \{ \left ( 0.72 + 0.44 \rm log
\frac{L}{L_{\star}} \right )
\frac{1.97x^{1.22}}{(x^2+0.78^2)^{1.43}} \right . \\ \left . + 1.6\ \rm exp [-0.4(L/L_{\star})]\frac{x^2}{x^2 + 1.5^2 \left ( \frac{L}{L_{\star}}\right )^{0.4}}  \right \} 
\rm km\ s^{-1}
\end{multline}

\noindent
where $x = R/R_{opt}$.  We consider models with $0.1 < L/L_{\star} < 5.0$, `optical' radii $R_{opt}$ (roughly three times the disk scale length, and defined by \citet{per96} to include $\sim 83\%$ of the light from an exponential disk; in practise our models trace the velocity field out to twice the optical radius)  ranging from 0.11-15\,pixels (with a plate scale of $\sim$0.42kpc/pixel at this redshift). We also allow the inclination and orientation of the rotation axis to vary within our previously determined constraints from harmonic decomposition and from a prior model using a rotating disk alone without simultaneously accounting for the tidal arm structures.  
For each pixel, the contribution from the two model components within a 21 by 21 pixel region equivalent to the size of the observational PSF are weighted according to their relative fluxes and convolved with the PSF.

Our best fit model is determined using a Levenberg-Marquardt least-squares minimization technique \citep{mark09}, and has $L = 0.25L_{\star}$, $R_{opt} = 10.7^{+3.0}_{-1.6}$ and an inclination angle of $47 \pm 12^{\circ}$. While the formal 1$\sigma$ parameter errors for our best fit model are 0.2\,pixels for $R_{opt}$ and $1.1^{\circ}$ for the inclination angle, we note that reasonable fits to the data can also be obtained with inclination angles ranging from $35^{\circ}$ to $59^{\circ}$ before gross mis-fitting of the data becomes clear in the model-subtracted residuals. As the assumed inclination angle has the greatest impact on the derived dynamical mass, the full parameter errors given above are derived after applying these limiting values for the inclination angle, and account for the loosest possible constraints on disk inclination for this source.  

\begin{figure*}
\epsscale{.80}
\includegraphics[scale=0.45]{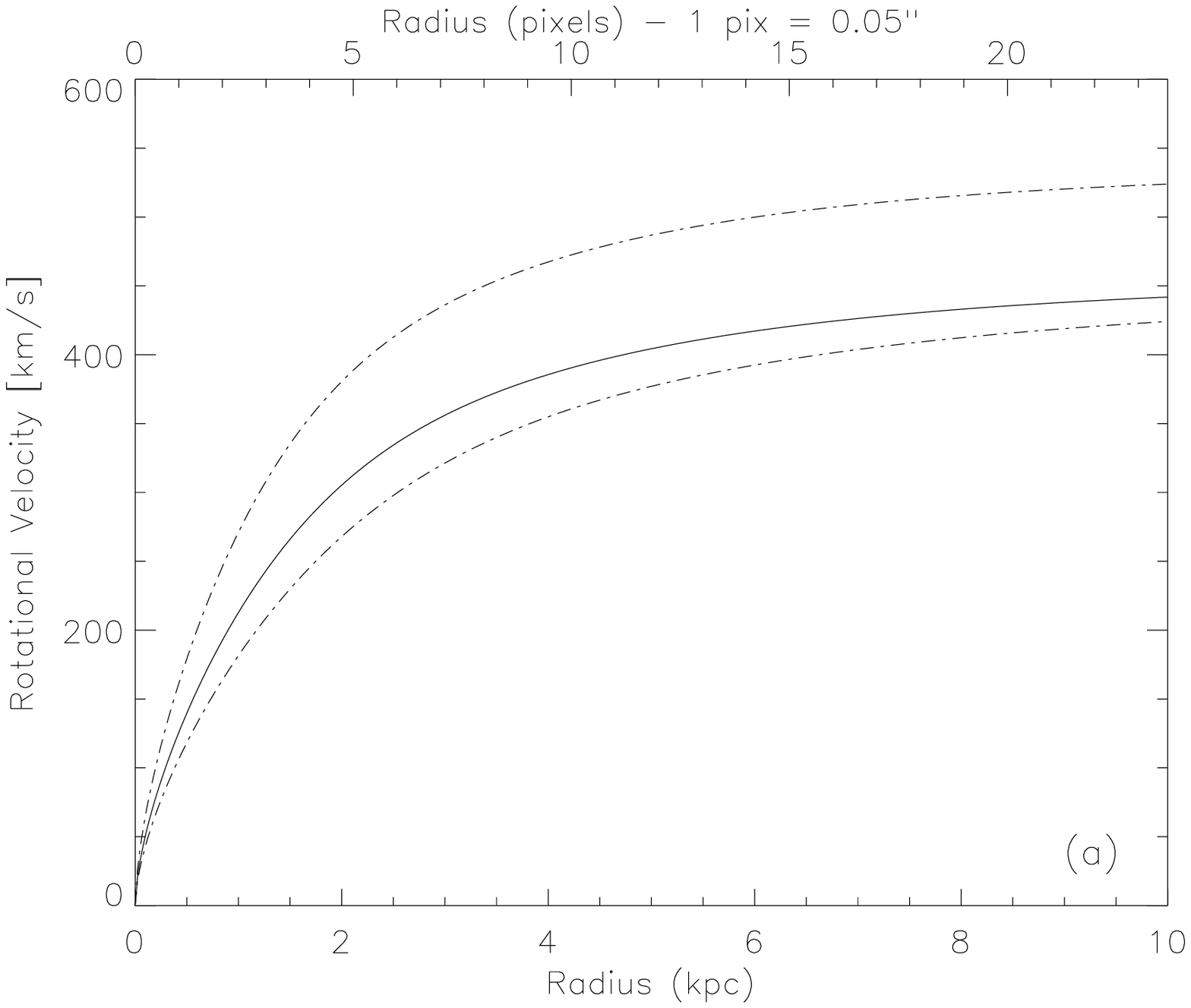}
\includegraphics[scale=0.45]{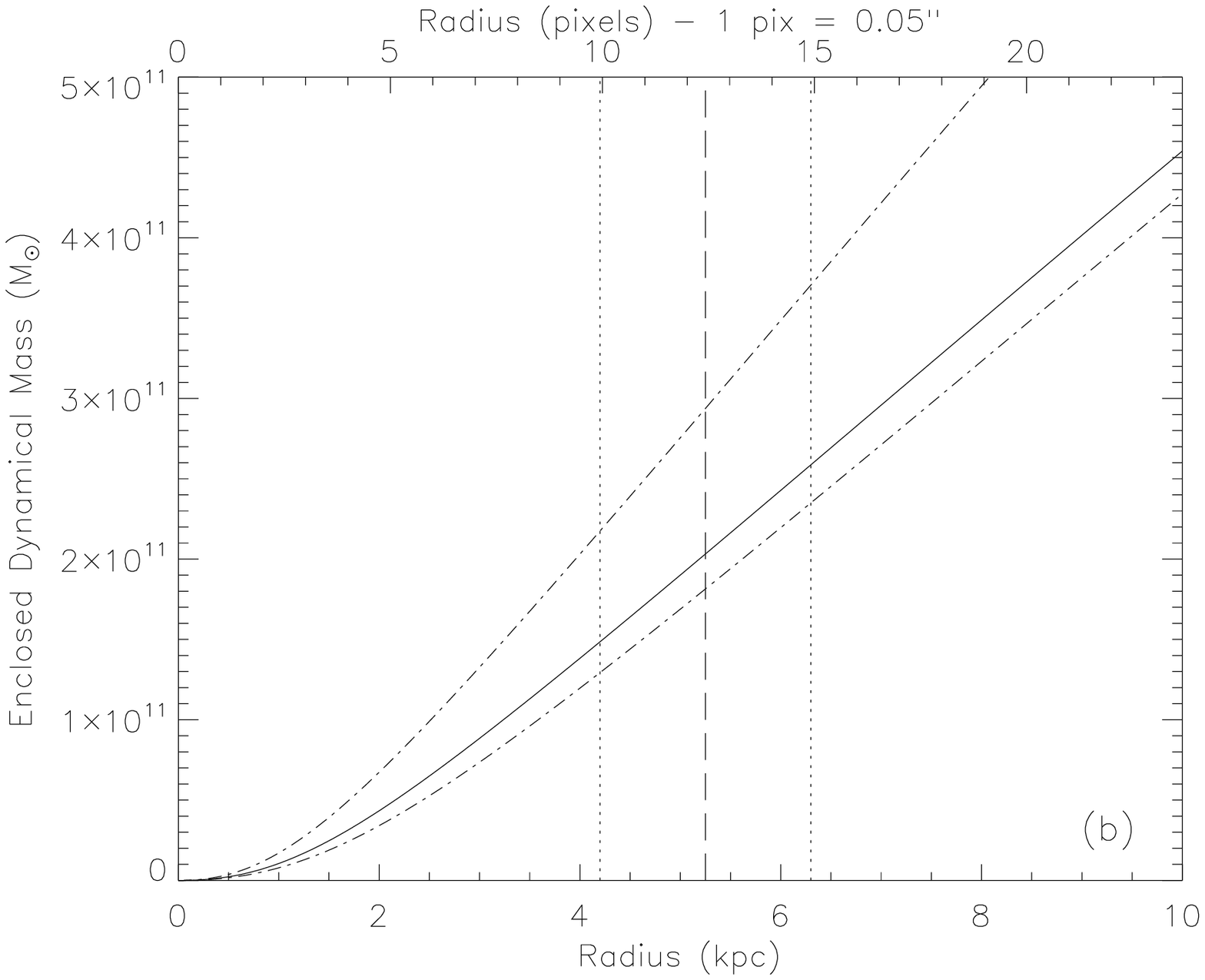}
\caption{Results of final velocity field modelling (see section 3.2.2) for the central rotating gas disk component which traces the galactic potential. (a -- left) Best fit rotation curve (solid line) and limiting rotation curves defined by the associated parameter errors (dot-dashed lines) for the model with $L/L_{star}=0.25$, $R_{opt} = 10.7^{+3.0}_{-1.6}$ pixels and inclination angle $47 \pm 12^{\circ}$. (b -- right) Enclosed dynamical mass, plotted as a function of radius, for the best fit model (solid lines), and within the associated parameter errors (dot-dashed lines). The vertical dotted lines define the region within which we can reliably trace the galactic potential, with the dashed vertical line defining the radius (5.25kpc) at which we determine our best-fit dynamical mass.\label{fig6}}
\end{figure*}

\begin{figure*}
\epsscale{.80}
\includegraphics[scale=0.27]{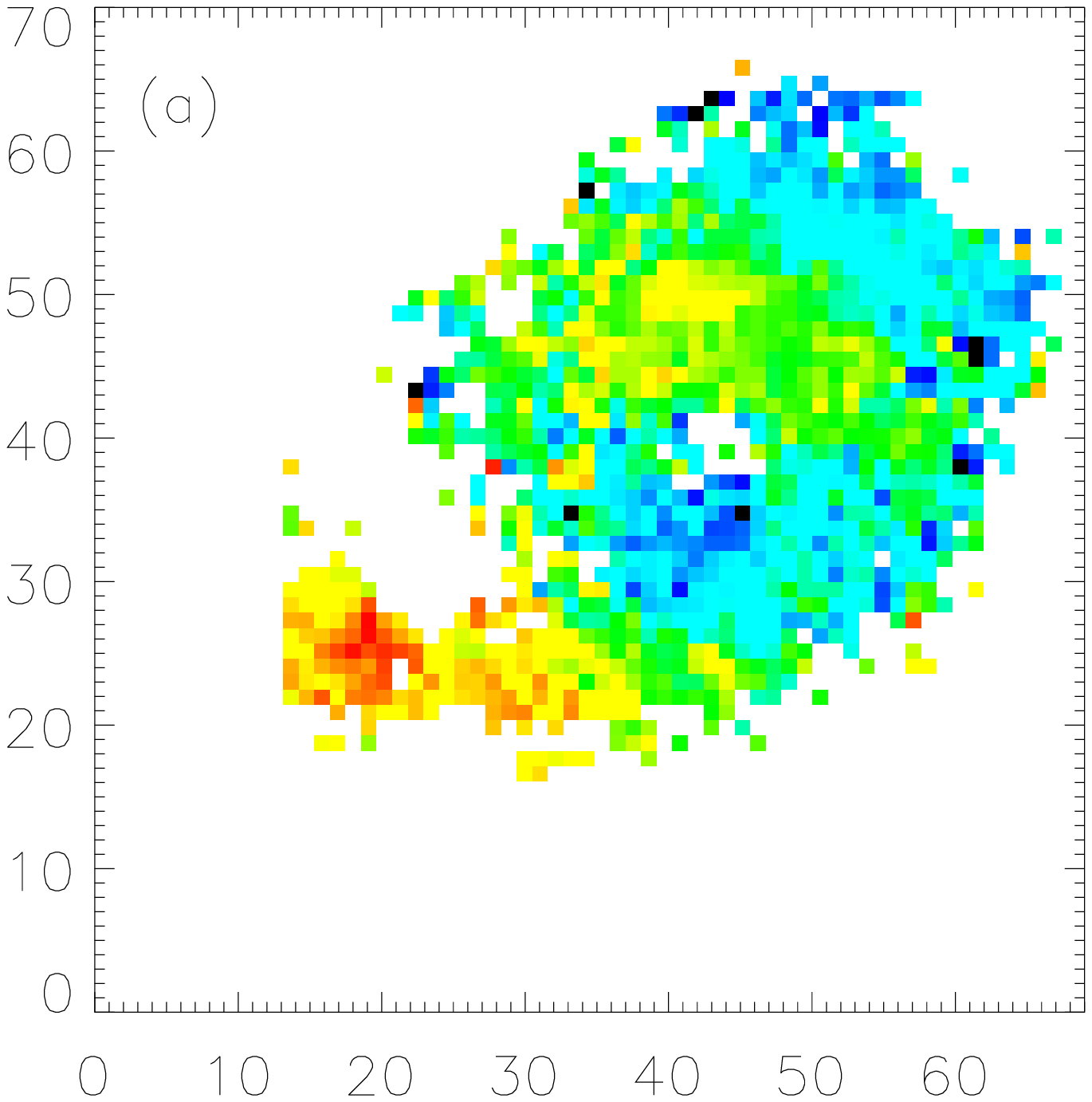}\hfill
\includegraphics[scale=0.27]{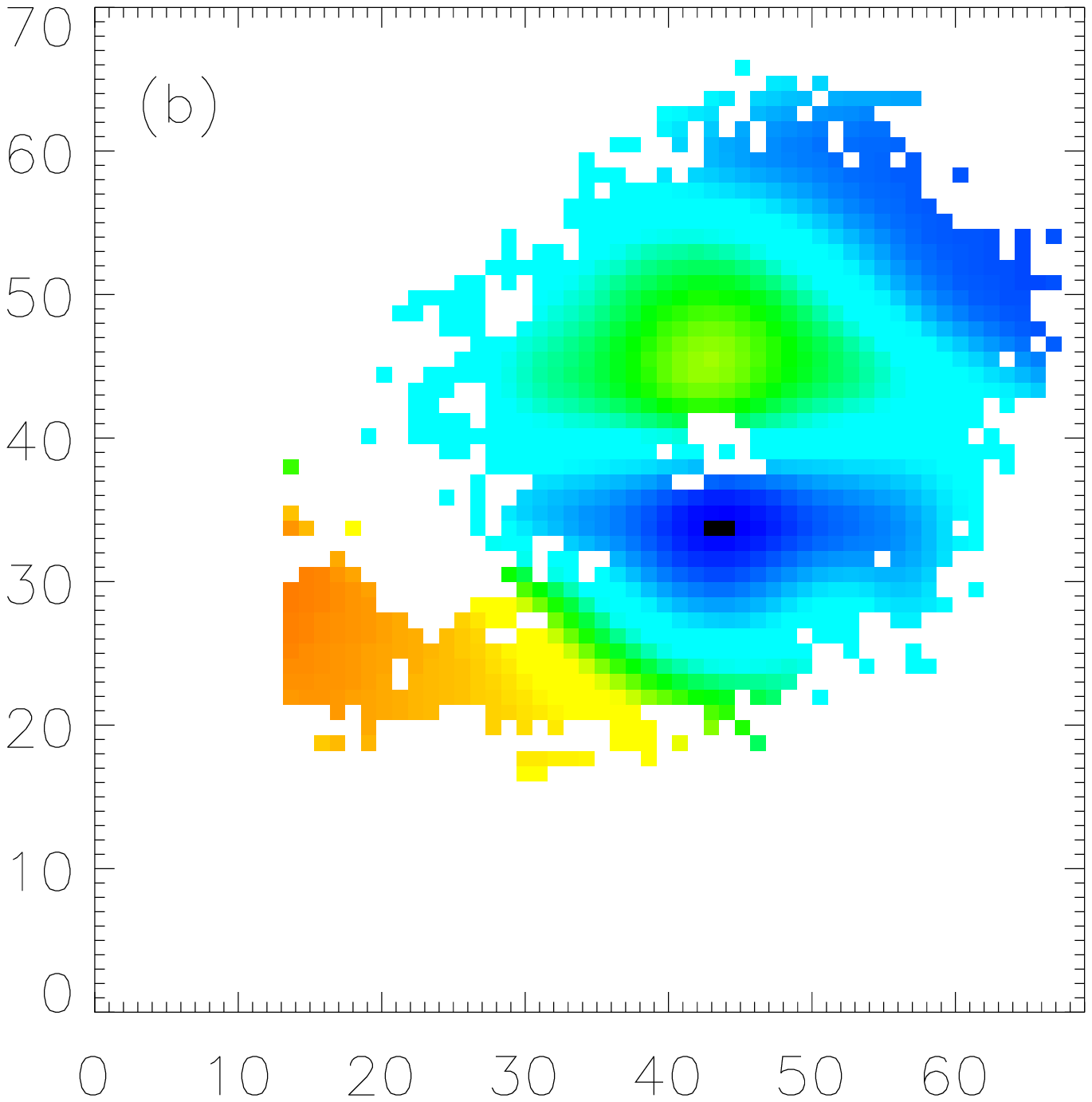}\hfill
\includegraphics[scale=0.27]{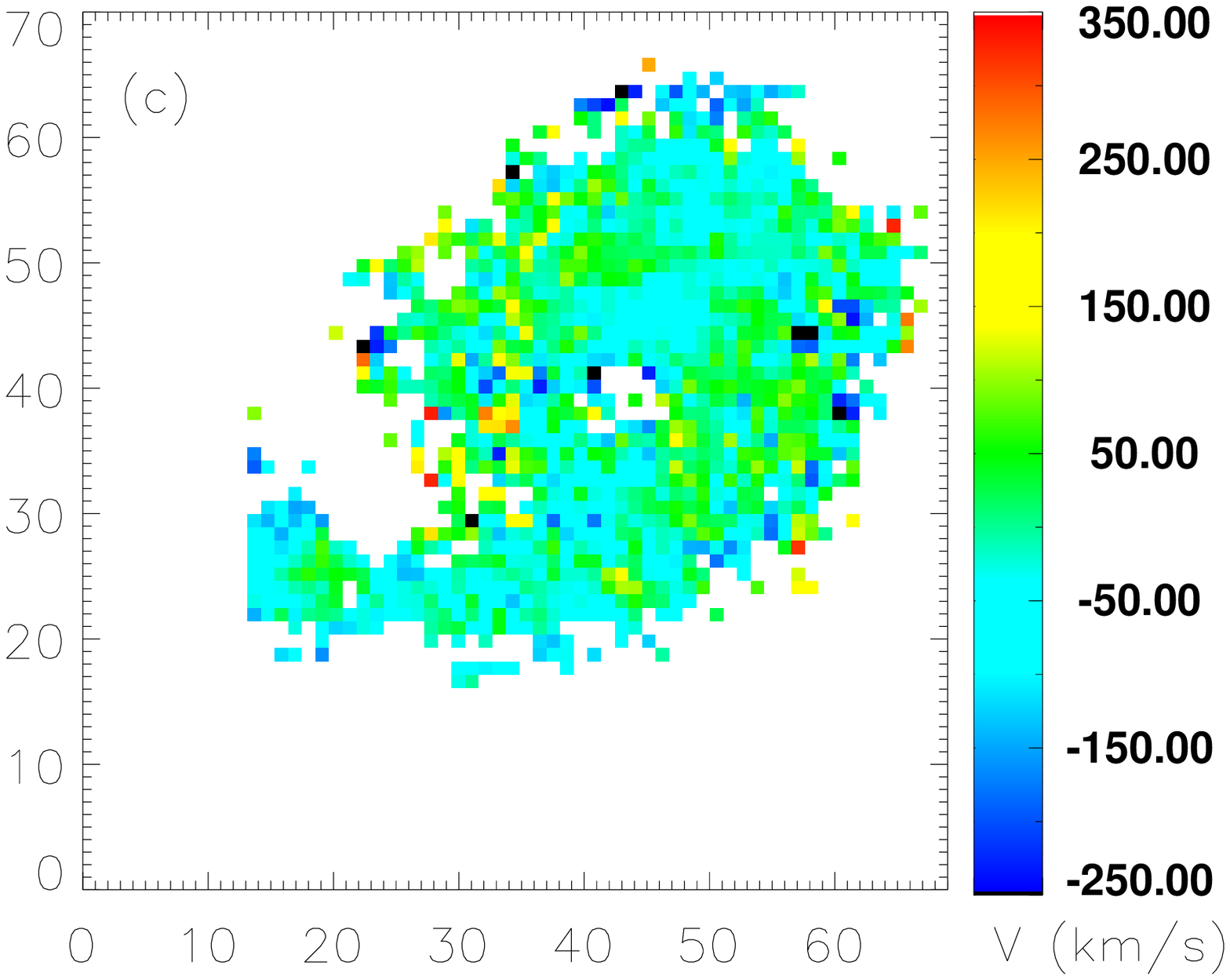}
\caption{Results of our velocity field modelling (see section 3.2.2). The masked data used in the model is displayed in frame (a), the best fit model in frame (b), and the model subtracted residuals in frame (c).\label{fig7}}
\end{figure*}

Fig.~\ref{fig6}a displays the de-projected rotation curve of the central gas disk component as a function of radius in the rest frame for our best-fit model. We also display deprojected rotation curves for our upper and lower limit mass models as defined by the parameter bounds above.  Derived from these curves using the gas motions as a tracer of the overall potential, we also plot the enclosed dynamical mass as a function of radius (Fig.~\ref{fig6}b).  This latter plot assumes that $v_{rot} = v_{circ}$. In reality, non-circular gravitational motions of the gas will also be present (e.g. streaming and turbulence in the ISM), and the derived dynamical mass profile is, strictly speaking, a lower limit. However, as is clear from Fig.~\ref{fig3}b, where the line emission originates solely from the centrally rotating regions -- the eastern and western quadrants of the centrally rotating structure, which are not subject to possible contamination by the tidal features -- the observed line widths are at their lowest and in fact are barely resolved above the instrumental resolution ($\sim 70\, \rm km\,s^{-1}$). The intrinsic line dispersion due to non-circular gravitational motion in the gas will be minimal, and thus $v_{rot} \approx v_{circ}$ is a good approximation. 
In Fig.~\ref{fig7}, we display the masked data, best fit model and model-subtracted residuals for the velocity field.

We can reliably trace the galaxy rotation curve out to a distance of between 10-15 pixels, equivalent to a galactic radius of 4.2 -- 6.3\,kpc. (Note that the 70\% encircled energy radius for the point spread function of our datacube is $\sim 0.2^{\prime\prime}$, equivalent to a projected distance of $\sim 1.7$kpc at the redshift of this source.) Ideally, we would determine a dynamical mass with reference to the scale length of the stellar light in the host galaxy, but no such information is available for this object. (While we have determined a best fit measure of $R_{opt} \sim 10.7$\,pixels for the {\it gaseous} emission for this galaxy, this does not necessarily have any bearing on the scale length of the stellar light, and should not be used as a proxy in this case.)  Within a radius of $5.25 \pm  1.05$\,kpc (i.e. 10 to 15 pixel radius), the enclosed dynamical mass of the host galaxy implied by our best fit model parameters (including errors) is $2.05^{+1.68}_{-0.74} \times 10^{11}\, \rm M_{\odot}$.

\subsection{Star formation in the host galaxy}

\begin{figure}
\epsscale{.80}
\includegraphics[scale=0.7]{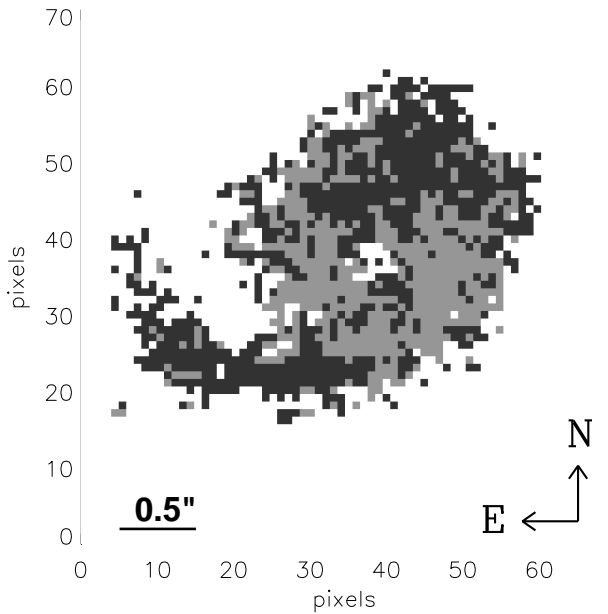}
\caption{Dominant ionization mechanism pixel flagging map. This plot illustrates in dark grey the regions where the emission line ratio log([N\textsc{ii}]6584/H$\alpha$) $< -0.5$, i.e. where the emission is likely dominated by stellar rather than AGN photoionization. The light grey pixels denote the regions for which AGN photoionization may instead be the dominant ionization mechanism. \label{fig8}}
\end{figure}

In addition to constraining the dynamical mass, the narrow H$\alpha$ emission can itself be used to estimate the star formation rate (SFR) in the host galaxy. We measure a total narrow H-alpha line luminosity of $1.22 \pm 0.21 \times 10^{43}\, \rm ergs\ s^{-1}$  from the unmasked regions of our datacube (including the 12\% uncertainty in the absolute flux calibration). A further maximum of 5\% of the total line flux may have been missed in the central twenty spaxels where errors due to the subtraction of QSO emission dominate, and from the small number of other masked pixels within the high signal-to-noise regions of narrow line emission, which we add to the upper error bound. The origin of the H$\alpha$ line emission is not necessarily solely due to the presence of young stars: AGN activity and shocks may also play an important role in the spatially extended ionization of the gas.  While we cannot place our data on a standard BPT diagram \citep{bpt, vo} without the addition of a second line ratio, sources with log([N\textsc{ii}]/H$\alpha$) $< -0.5$ are predominantly observed to be star-forming rather than AGN-photoionized sources \citep[e.g. ][]{groves06}. Fig.~\ref{fig8} displays the unmasked regions of our datacube shaded according to whether the line ratio  log([N\textsc{ii}]/H$\alpha$) is $< -0.5$ or $> -0.5$, i.e. whether or not stellar photoionization is likely to be the dominant ionization mechanism, or whether the observed line emission is better  considered as part of the extended
 narrow line region of the AGN. Star formation dominates in the tidal features, but is equally important in some of the more luminous regions of H$\alpha$ emission towards the center of the galaxy.
 If the H$\alpha$ line emission is due solely to star formation activity alone, the observed line flux is equivalent to a star formation rate of $96 ^{+22}_{-17}\, \rm M_{\odot}\, \rm yr^{-1}$ using the scaling of \citet{kenn98}. As the H$\alpha$ emission can be produced by AGN photoionization as well as via photoionization by young stars, this value represents a robust upper limit on the overall star formation in this system. In Fig.~\ref{fig9}a, we display this cumulative SFR as a function of radius.  In Fig.~\ref{fig9}b, we display the cumulative SFR as a function of radius considering only the data with line ratio values of log([N\textsc{ii}]/H$\alpha$) $< -0.5$.  In this case, we observe an overall SFR of $50 \pm 10 \rm M_{\odot} \rm yr^{-1}$, derived from H$\alpha$ emission that is {\it unlikely} to have been photoionized by shocks or by the AGN, providing a useful lower limit on this quantity. The average integrated SFR per square kiloparsec (projected) is displayed for all data and also for the data with log([N\textsc{ii}]/H$\alpha$) $< -0.5$ in Fig.~\ref{fig9}c and Fig.~\ref{fig9}d respectively. These plots clearly illustrate the importance of star formation throughout this galaxy, and that it is not restricted solely to either the inner or outer regions.
     
\subsection{Broad line emission and the QSO properties}

Taking the literature data for J090543.56+043347.3, the mass of the central black hole has been estimated as $9.02 \pm 1.43 \times 10^8 M_{\odot}$, based on the width of its  Mg\textsc{ii} emission line and the local continuum luminosity \citep{shen08}.  Additionally, this source has a bolometric luminosity of $3.10 \times 10^{39} \rm W$ \citep{shen08}, corresponding to a factor of 0.25 $L_{EDD}$ for a black hole of this mass.

We can also determine a black hole mass based on measurements of the broad H$\alpha$ emission line. Using the formulism of \citet{gh05},

\begin{multline}
M_{BH} = (2.0^{+0.4}_{-0.3}) \times 10^6 \left (
\frac{L_{H\alpha}}{10^{42} \rm ergs\, s^{-1}} \right )^{0.55 \pm
  0.02}\\ \times \left ( \frac{FWHM_{H\alpha}}{10^3 \rm km\, s^{-1}} \right )^{2.06 \pm 0.06} M_{\odot}.
\end{multline}

We derive a spatially integrated QSO spectrum by combining the spectrum of the peak spaxel of our flux-calibrated and continuum subtracted datacube (dominated by QSO emission) with the PSF profile derived from the broad line flux across the field of view. The broad H$\alpha$ line has a measured FWHM of $3160 \pm 170 \rm km \, s^{-1}$, and an overall luminosity of $109 \pm 13 \times 10^{42} \rm ergs\, s^{-1}$, resulting in a black hole mass estimate of $2.83^{+1.93}_{-1.13} \times  10^8 M_{\odot}$.  We note that the Mg\textsc{ii} and H$\alpha$ based black hole masses differ by a factor of $\sim 3$, and that the observed bolometric luminosity of \citet{shen08} is equivalent to a factor of $0.79 L_{Edd}$ if we assume the H$\alpha$-derived black hole mass, which is likely to be subject to a slightly lower overall systematic uncertainty.

\section{Discussion and Conclusions}

The driving science question behind this work is to investigate the  black hole mass  vs. host galaxy  mass relation at high redshifts, in a direct and unbiased manner as possible.  Obtaining dynamical quasar host galaxy masses via narrow line emission kinematics is an ideal way forward.  Although the existence of complex velocity structures can potentially present difficulties for such a method, our pilot object J090543.56+043347.3 despite showing a non-simple velocity field can nonetheless be successfully modelled.
Having obtained a measure of both black hole and host galaxy dynamical mass for J090543.56+043347.3, in Fig.~\ref{fig10} we contrast it with the canonical low redshift $\rm M_{BH}$ vs.\ $\rm M_{Bulge,Dyn}$ relation of \citet{hr04} and also the more recent $\rm M_{BH}$ vs.\ $\rm M_{Stellar}$ and $\rm M_{BH}$ vs.\ $\rm M_{Bulge,Dyn}$ relations of \citet{sani}, based on {\it Spitzer}/IRAC data.  The dynamical bulge masses of \citet{hr04} plotted on Fig.~\ref{fig10}a are either derived via first confirming that the light profiles of the galaxies are bulge-dominated, and then solving the spherical Jeans equation, or taken directly from the literature. This sample uses reliable pre-existing black hole mass estimates from the literature, derived from gas and stellar kinematics and primarily sourced from \citet{tre02}. For the points on Fig.~\ref{fig10}b, \citet{sani} carry out 2-d image decomposition of the 3.6$\mu\ m$ data for their sample, and then derive virial dynamical galaxy masses as $\rm M_{Bulge,Dyn} = 5 \sigma^2 R_e /G$. Once again, the black hole masses for this sample are reliable estimates from the literature based on stellar or gas dynamics, or masers.  
Note that there is no significant difference between the regression lines for $\rm M_{BH}$ vs.\ $\rm M_{Bulge,Dyn}$ and $\rm M_{BH}$ vs. $\rm M_{Stellar}$ derived by \citet{sani} from the 3.6$\mu\,$ mass-to-light ratio including colour-correction terms. Two separate points are plotted for the different Mg\textsc{ii} and H$\alpha$ based black hole masses of J090543.56+043347.3. We find that, using the Mg\textsc{ii} based black hole mass, the position of this object on either plot is a factor of $\sim 2$ from the mean $z=0$ relation, but well within the scatter. Using our H$\alpha$ based black hole mass measurement, the position of this object lies directly upon the mean $z=0$ relation. This no-evolution result is also consistent with the theoretical modelling of \citet{jahn11}, in which the correlation can be fully explained to have emerged at high redshifts from hierarchical assembly and merging statistics. Due to the drop both in merger rate as well as SF and BH accretion rate densities in the universe since $z\sim2$, their explanation predicts the overall scaling relations to be largely in place by $z=1$.

It should be noted that on Fig.~\ref{fig10} we compare the dynamical mass derived for a z=1.3 galaxy with an unknown stellar morphology and a rotating gas disk with stellar masses and bulge dynamical masses in the {\it local} universe. Given this, it is therefore worthwhile to consider the likely subsequent evolution of this system over the intervening 8-9 Gyrs.  The major mechanisms for the growth and evolution of the black hole and its host galaxy are BH accretion, star formation, merger activity and disk-to-bulge reprocessing, which we now consider in turn.

The bolometric quasar luminosity of J090543.56+043347.3 \citep{shen08} is equivalent to a black hole accretion rate of $\sim5.4\, \rm M_{\odot} \rm yr^{-1}$ (assuming mass is converted to luminosity with an efficiency of 10\%). Quasar lifetimes are typically between $10^7$ and $10^8$ years \citep{yt02, mar04, hh09, kell10}. Taking the average of the two black hole mass measurements for J090543.56+043347.3 ($\sim 6 \times 10^8 \rm M_{\odot}$) and a timescale of $5 \times 10^7$ years, an accretion rate of $\sim5.4\rm M_{\odot} \rm yr^{-1}$ would result in the central black hole growing by $\sim 45\%$  in total over the current active cycle of AGN-mode accretion. The absolute maximum possible growth factor for this black hole (assuming: [i] that it is observed at the very {\it start} of its current activity cycle, [ii] a maximum accretion lifetime of $10^8$ years, and [iii] the lower of the two black hole mass estimates) is a factor of $\leq 2$, though we do stress that growth approaching this limiting value is unlikely.
 At the same time, it is statistically not very likely that this BH would undergo any further episodes of AGN activity between $z=1.3$ and $z=0$ \citep[given the expected AGN fraction as a function of redshift;][]{bluck}.

\begin{figure*}
\epsscale{1.0}
\includegraphics[scale=0.45]{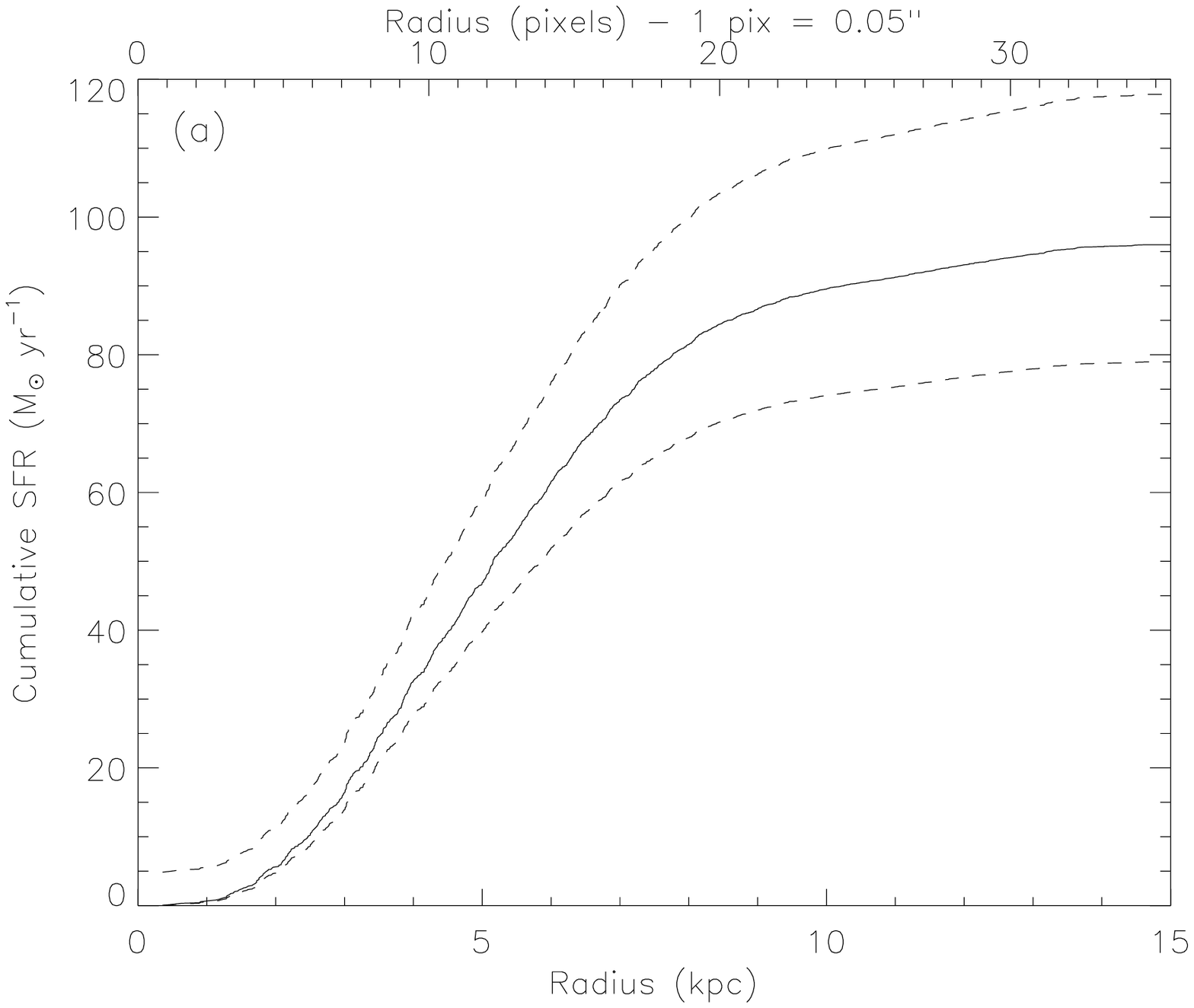}\hfill
\includegraphics[scale=0.45]{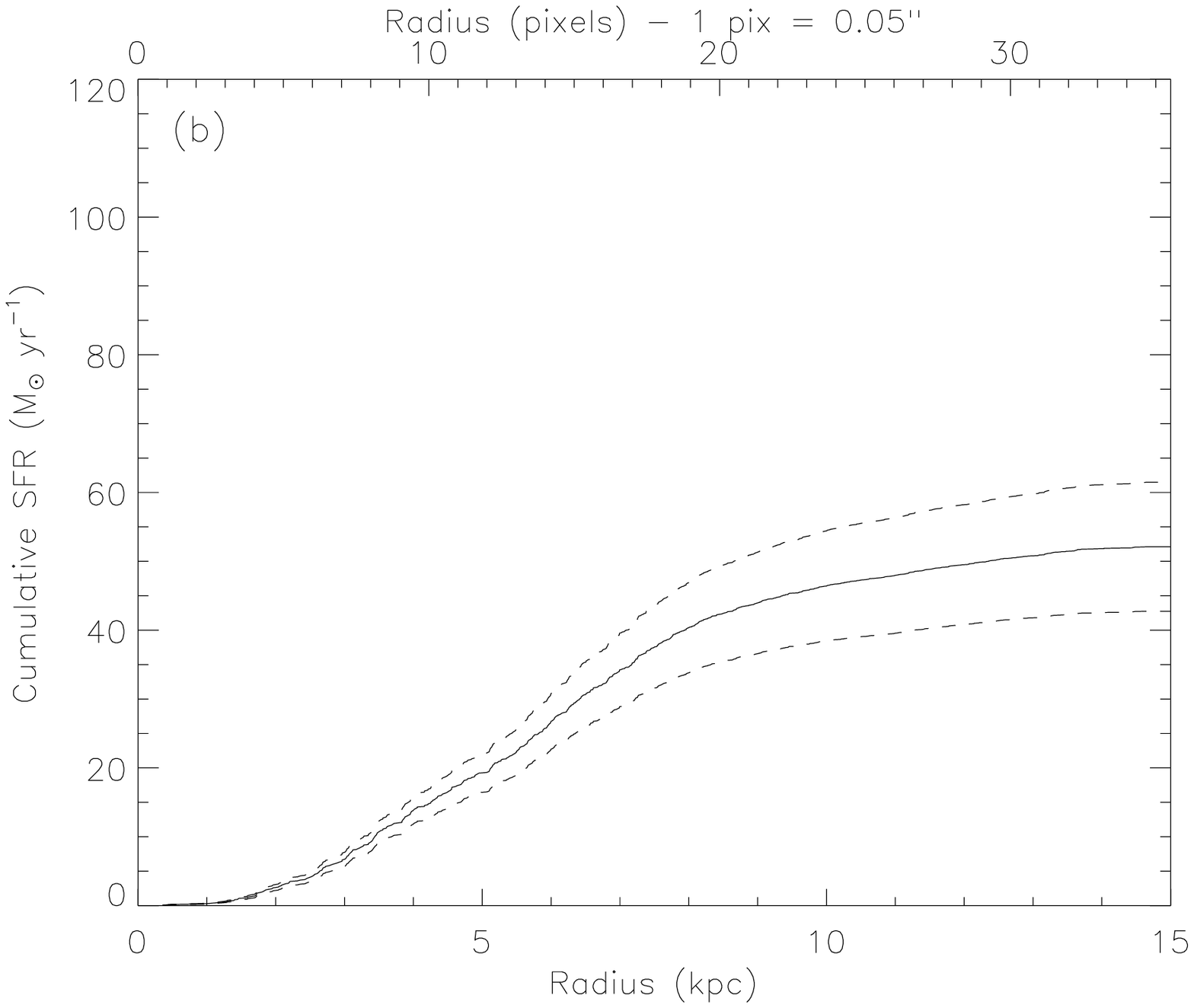}\\
\includegraphics[scale=0.45]{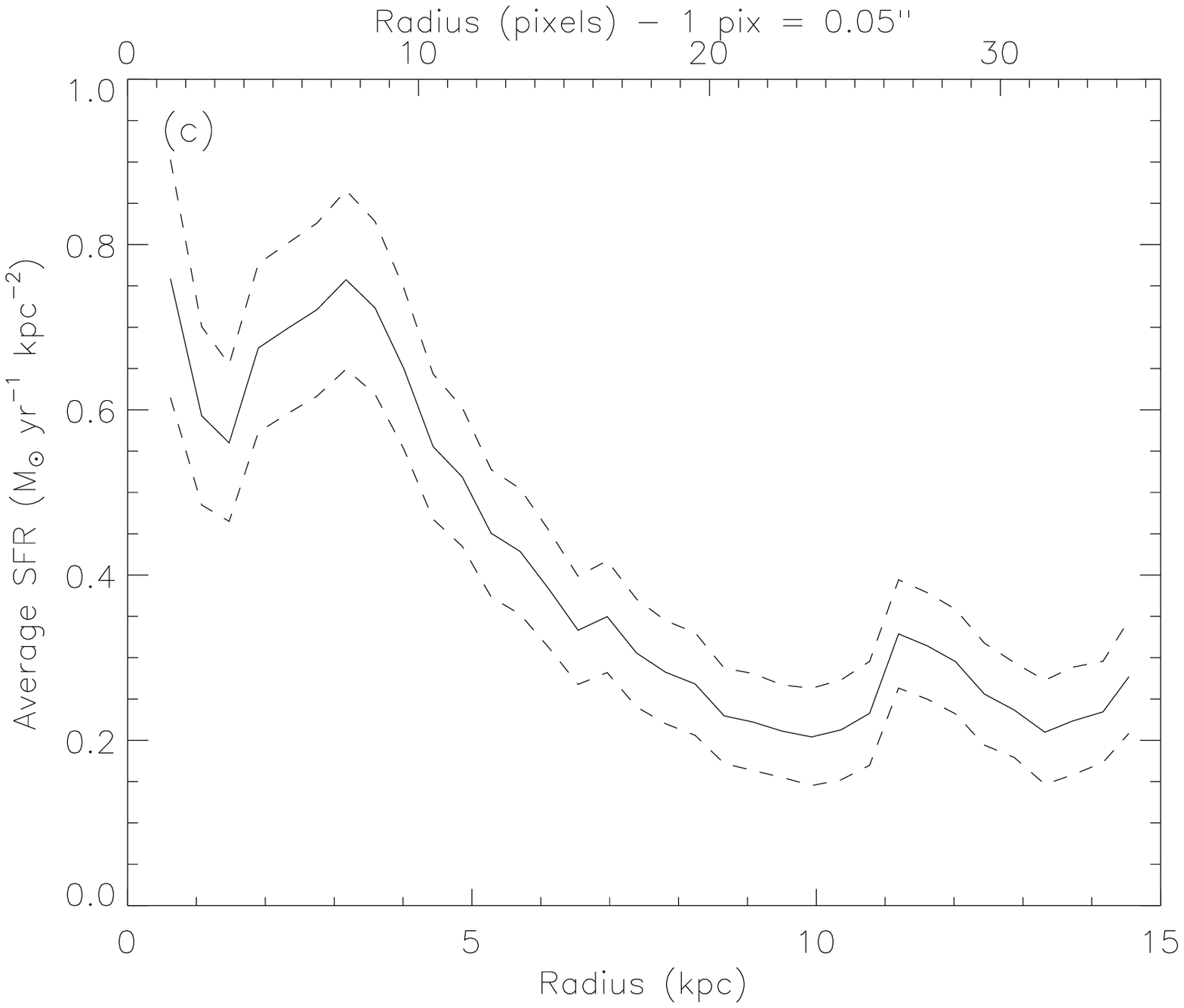}\hfill
\includegraphics[scale=0.45]{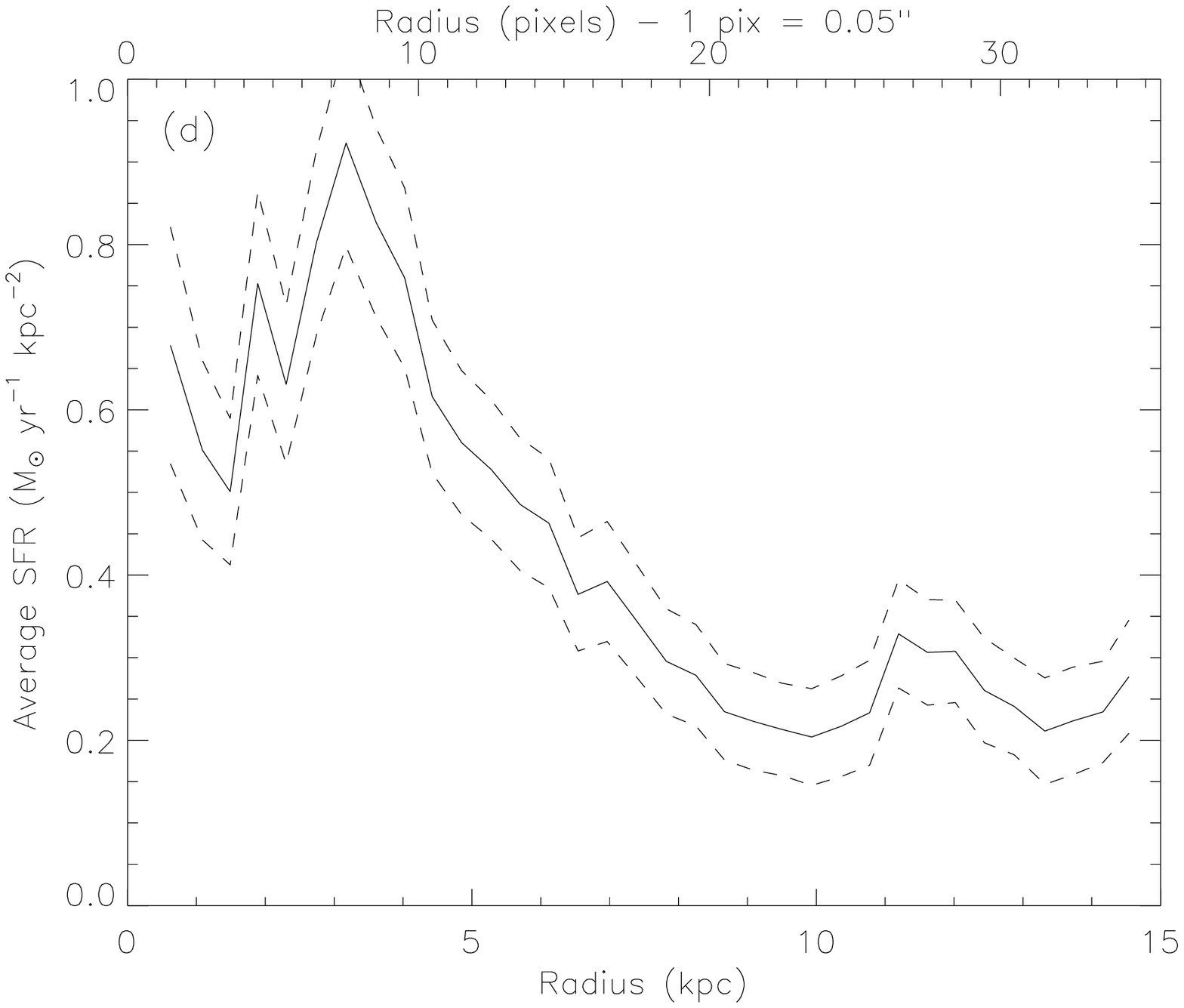}
\caption{(a -- top left) Cumulative SFR (solid line) plus error bounds (dashed lines) as a function of radius, calculated using the scaling of \citet{kenn98}. (b -- top right) As frame (a), but only including data from spaxels with an emission line ratio log([N\textsc{ii}]6584/H$\alpha$) $< -0.5$, i.e. where the emission is likely dominated by stellar rather than AGN photoionization (see fig~8.). (c -- lower left) Average SFR per square kiloparsec (projected) as a function of radius (solid lines), plus error bounds (dashed lines). (d -- lower right) As frame (c), but only including data with   log([N\textsc{ii}]6584/H$\alpha$) $< -0.5$.  \label{fig9}}
\end{figure*}

Considering star formation in the host galaxy, our observed H$\alpha$ flux is consistent with a SFR of up to 100\, $\rm M_{\odot} \rm yr^{-1}$ (section 3.3). Only accounting for the line emission which is likely due to ionization by young stars rather than  the AGN or shocks  (particularly true of the tidal arms and the areas of high emission line intensity within the host galaxy: see Fig.~\ref{fig3}d, Fig.~\ref{fig9}) gives a SFR of 40-60$\rm M_{\odot} \rm yr^{-1}$. The observed star formation rate is significantly higher than average \citep[e.g. ][]{karim} for a galaxy at this redshift with a stellar mass approximating our derived dynamical mass ($2.05^{+1.68}_{-0.74} \times 10^{11}\, \rm M_{\odot}$ within a radius of $5.25 \pm  1.05$\,kpc). But this may not be surprising given the clear presence of highly star-forming tidal features (Fig.~\ref{fig3}). 

As an aside, we note that triggering of the quasar activity via interactions/mergers is a possibility for this particular AGN, given the clearly discernable merger signature in the gas. The high projected star formation rate per square kiloparsec towards the center of the host galaxy (within a radius of 4kpc; see Fig.~\ref{fig9}c and Fig.~\ref{fig9}d) might also be suggestive of links between mergers, AGN activity and circum-nuclear star formation in the case of this source, or of a nuclear star formation ring such as that observed in the case of NGC 1097 \citep{vdven2010}.  However, the triggering of AGN activity via mergers remains unproven and we are far from being able to e.g.\ determine and compare the age of the AGN\footnote{The only limit we
have is a mininum age of the AGN of $\sim2\times10^4$ years, from the fact
that we see gas ionized by the AGN at 0\farcs7 distance.} with that of
stars newly formed by a merger. Overall, there is very little direct observational evidence 
that galaxy mergers trigger AGN even in individual cases at lower
redshifts \citep[e.g.][]{cana00}, while on a population basis major
merging can clearly be ruled out as the dominating mechanism at $z<1$
\citep{cist11}.

Assuming that the
local scaling relations for black hole mass vs. galaxy stellar mass were already 
approximately in place at $z \sim 1.3$, we note that mergers involving galaxies with similar $\rm M_\mathrm{BH}$/$\rm M_\mathrm{Total}$ ratios  will (statistically) lead to the same proportional growth for both black hole and host galaxy, leaving the relation unchanged.
It is expected that a stereotypical galaxy with the same mass as that determined for J090543.56+043347.3 would undergo at most one more major merger event \citep{aday10}, and quite likely be restricted to more minor accretion events. Therefore, both the host galaxy and its central BH are likely to grow by a maximum of approximately 50\% in mass by the present day due to merger activity alone \citep[see, e.g. ][]{hopk10, vd10}. Similarly, under the assumption that the observed scaling relations for $\rm M_\mathrm{BH}$ vs.\ $\rm M_\mathrm{Host}$ are already in place at $z=1.3$, merger activity alone would shift the exact placement of this object on Fig.~\ref{fig10} in a direction roughly parallel to the observed mean $z=0$ scaling relation.

\begin{figure}
\includegraphics[width=\columnwidth,bb=35 12 440 329]{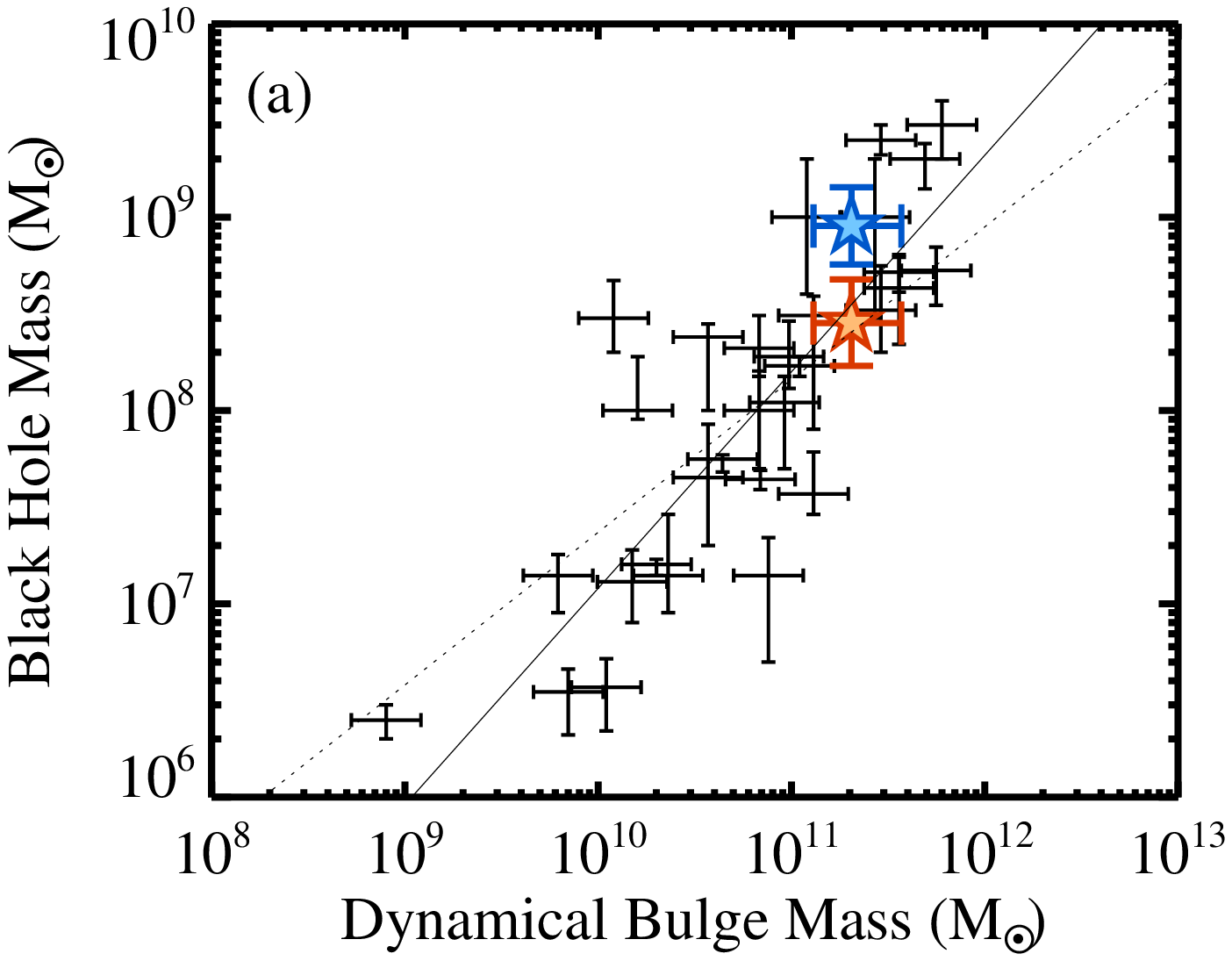}\\
\includegraphics[width=\columnwidth,bb=35 12 440 329]{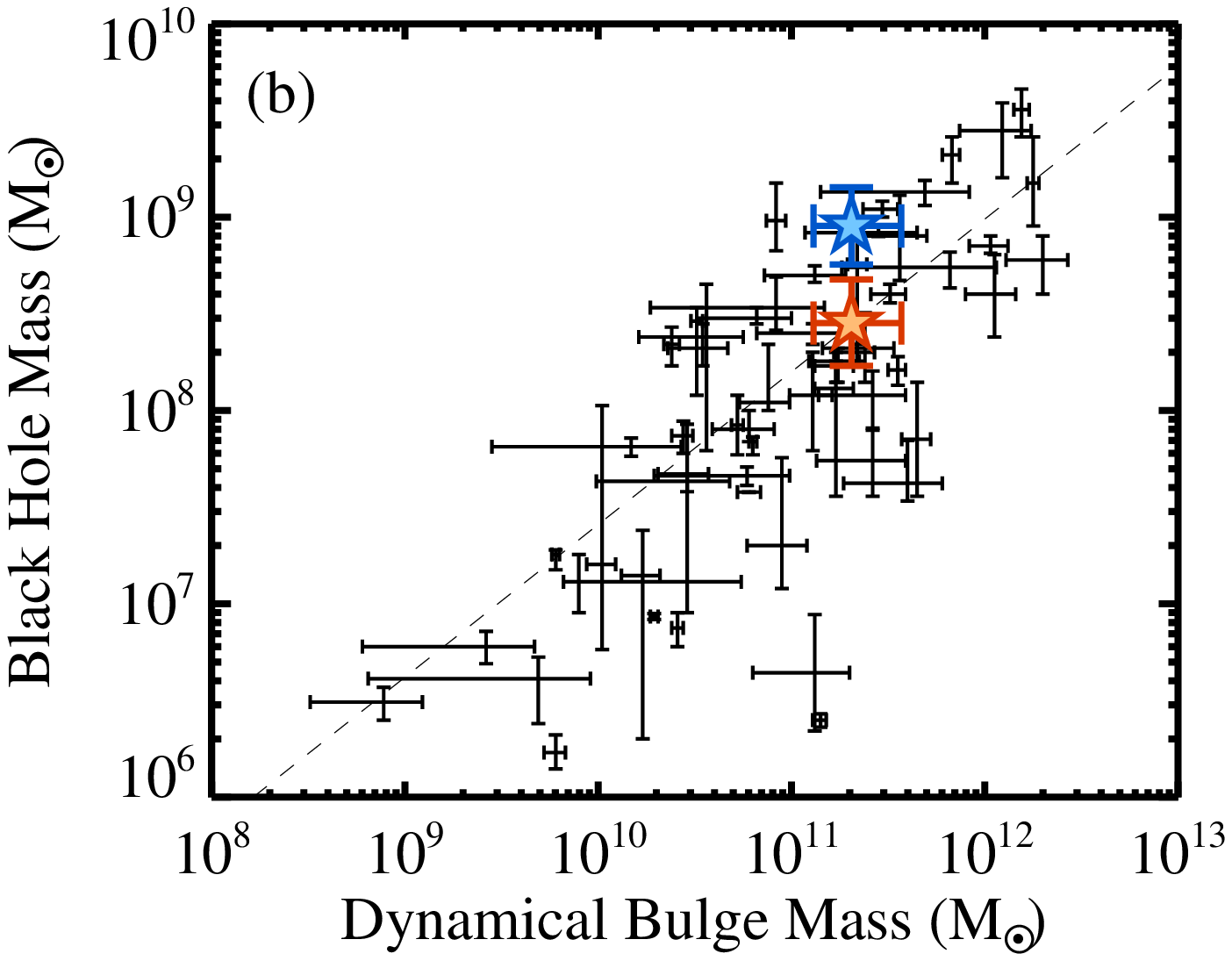}
\caption{J090543.56+043347.3 and the black hole mass vs. host galaxy mass relation (a -- top) Position of J090543.56+043347.3 (stars) relative to the $z=0$ $\rm M_{BH}$ vs. $\rm M_{Bulge,Dyn}$ correlation, with datapoints from \citet{hr04}. The upper (blue) star uses the black hole mass derived from the Mg\textsc{ii} emisison line, while the lower (red) star uses our H$\alpha$ based black hole mass. Note that the total dynamical mass is plotted for J090543.56+043347.3, and that this represents an upper limit to the bulge mass at $z=1.3$. The solid line displays the bisector linear regression fit  of \citet{hr04}, while the dotted line gives the linear regression fit for the $\rm M_{BH}$ vs. $\rm M_{stellar}$ relation derived by \citet{sani}. (b -- bottom) Position of J090543.56+043347.3 (stars, color-coded as in frame (a)) relative to the $z=0$ $\rm M_{BH}$ vs. $\rm M_{Bulge,Dyn}$ correlation, with datapoints and linear regression fit (dashed line -- note that this is indistinguishable from the $\rm M_{BH}$ vs. $\rm M_{stellar}$ regression fit denoted by the dotted line on frame (a)) from \citet{sani}. \label{fig10}}
\end{figure}

Even though the gas and stars trace the same gravitational potential, the gas and stellar kinematics can be expected to differ. While the observed gas kinematics for the bulk of this system are relatively settled and consistent with a rotating disk, the stellar morphology can not be inferred from the gas morphology, and we emphasize that it remains unknown.  However, given the observational evidence for ongoing tidal activity, even if this galaxy is restricted in the future solely to minor accretion rather than major merger events,  the observed dynamical mass at z=1.3 can be readily reprocessed into a stellar bulge. This object is almost certainly the progenitor of a $z=0$ bulge dominated galaxy.

What bearing does the combination of all our observations have on the likely subsequent evolution of this system and on its position relative to the local scaling relations?

On the basis of the observations of J090543.56+043347.3, we can quantify the relative rate at which both the mass of the central black hole and of the host galaxy's stellar population are growing at the epoch at which we observe them.  The annual increase in black hole mass of $5.4 \rm M_{\odot} yr^{-1}$, as derived from the quasar's bolometric luminosity, is just ten times less than the observed (conservative) star formation rate of $\sim 50 \rm M_{\odot} yr^{-1}$. By contrast, the scaling relations predict a black hole which is of the order $\sim 500$ times less massive than its host galaxy. While the central black hole is actively accreting, the host galaxy is forming new stars at a much more moderate rate relative to its overall {\it stellar} mass. 

Of course, we know that the duty cycle of star formation for $z \sim 1.5$ galaxies is far higher than the duty cycle of active black hole accretion.  Over a single AGN activity cycle lasting typically $10^7$ to $10^8$ years, the proportional increase in mass of the black hole is expected to be $\sim 45\%$, as noted above.  A similar proportional increase in the stellar mass of the host galaxy can be achieved if the observed level of star formation persists over a substantially longer timescale than the AGN activity: a galaxy of $2 \times 10^{11} \rm M_{\odot}$ with a star formation rate of $\sim 50 \rm M_{\odot} yr^{-1}$ would increase in stellar mass by 45\% over the course of $\sim 2$Gyr.  In such circumstances, the black hole mass vs. host galaxy mass ratio of this object would remain consistent with the local relation. However, we note that the required star formation timescale of several Gyr is well beyond the typical merger timescales. 

An alternate approach to constraining the expected future increase in stellar mass of J090543.56+043347.3 with time until $z=0$ is to consider star formation rates for the galaxy population as a whole.
More specifically, following the cosmic average \citep[e.g. ][]{karim} the expected levels of star formation activity would result in a growth of typically a factor of about 1.5 to 3 in stellar mass between $z=1.3$ and the present day, depending on the exact details of the subsequent star formation activity within the galaxy. 

We can combine this empirical estimate of the stellar mass growth through star formation activity with the increase in mass due purely to the expected merger activity \citep[$\sim 50\%$; ][]{hopk10}. Overall, the host galaxy is very likely to more than double in mass (increasing by $\ga 0.7\,$dex in $\rm log(M_{total})$) between $z=1.3$ and $z=0$. As the majority of any disk stellar mass will have been re-processed into the galactic bulge via galaxy mergers or tidal interactions, we would expect $\rm log(M_{total,z=0}) \approx \rm log(M_{Bulge,z=0})$.

Similarly, we can combine the expected growth in BH mass due to AGN accretion and merger activity to predict the overall increase in $\rm M_{BH}$ by $z=0$.  These processes are likely to result in an overall increase of $\la 0.25\,$dex in $\rm log(M_{BH})$ for a system such as this; taking our less-likely absolute upper limit for BH accretion, the maximum increase in $\rm M_{BH}$ is $\la 0.7\,$dex. 

Any change in position of this object by $z=0$ relative to the local relation will therefore most likely be dominated by changes in the host galaxy mass. 
Overall, regardless of the exact details of the further evolution of this object, it is most likely that the average of the two points plotted for this object on Fig.~\ref{fig10} will move to a position which is even {\it closer} to the best fit z=0 relations displayed in Fig.~\ref{fig10}, and in any case will remain consistent with the observed scatter.

In conclusion, we return to the observational aspects of the work presented here. We believe that we have presented a study that is ground-breaking in the respect that it demonstrated  how AO-assisted near-IR integral field spectroscopy can be successfully applied, and used to determine the dynamical mass profile for the host galaxy of a typical quasar. The observed BH masses and overall dynamical mass of the
host galaxy of this object at $z=1.3$ are consistent with the observed present-day 
relations for $\rm M_\mathrm{BH}$ vs.\ $\rm M_\mathrm{Bulge,Dyn}$ and $\rm M_\mathrm{BH}$ vs.\ $\rm M_\mathrm{Stellar}$, and aside from a potential conversion
of disk- to bulge-mass through minor mergers or the likely ongoing major merger,
no significant evolution in the black hole vs. stellar mass relationships between $z=1.3$ and $z=0$ is required or expected.
Provided that one takes into account the minimum extended emission line flux requirements and the likelihood of favourable gas kinematics and orientation in the observed velocity field, this particular diagnostic route is widely applicable to other quasars in the redshift range $1 < z < 3$, at the peak epoch of quasar activity. Analysis of a further seven objects in our pilot sample is currently underway. Although J090543.56+043347.3 itself displays a relatively impressive star formation rate, this diagnostic method is clearly feasible at much lower narrow H$\alpha$ fluxes, while alternatives such as CO-based dynamical masses are most successful for extreme, atypical objects.

\acknowledgements{
KJI and KJ are funded through the Emmy Noether Programme of the German Science Foundation (DFG). We thank Ric Davies, Chien Peng and Arjen van de Wel for very useful discussions, Eleonora Sani for providing the $z \sim 0$ data which appears in Fig.~\ref{fig10}b, and the anonymous referee for their detailed consideration of the manuscript.

{\it Facilities:} \facility{VLT:Yepun}.

}


\begin{thebibliography}{59}

\bibitem[{{Baldwin, Phillips \& Terlevich}(1981){Baldwin}{Phillips}{Terlevich}}]{bpt}
{Baldwin}, J.~A., {Phillips}, M.~M., {Terlevich}, R., 1981, \pasp, 93, 5

\bibitem[{{Becker, White \& Helfand}(1995){Becker}{White}{Helfand}}]{first1}
{Becker}, R.~H., {White}, R.~L., {Helfand}, D.~J., 1995, \apj, 450, 559


\bibitem[{{Bennert} {et~al.}(2010){Bennert},{Treu},{Woo},{Malkan},{Le Bris},{Auger},{Gallagher},{Blandford}}]{benn10}
{Bennert}, V.~N., {Treu}, T., {Woo}, J.-H., {Malkan}, M.~A., {Le Bris}, A., {Auger}, M.~W., {Gallagher}, S., {Blandford}, R.~D., 2010, \apj, 708, 1507

\bibitem[{{Bluck} {et~al.}(2011){Bluck}, {Conselice}, {Almaini}, {Laird}, {Nandra}, {Gr\"utzbauch}}]{bluck}
{Bluck}, A.~F.~L., {Conselice}, C.~J., {Almaini}, O., {Laird}, E.~S., {Nandra}, K., {Gr\"utzbauch}, R., 2011, \mnras, 410, 1174

\bibitem[{{Bonnet} {et~al.}(2004){Bonnet+}}]{bon04}
{Bonnet}, H., et~al., 2004, Messenger, 117, 17

\bibitem[{{Borys} {et~al.}(2005){Borys}{Smail}{Chapman}{Blain}{Alexander}{Ivison}}]{bory05}
{Borys}, C., {Smail}, I., {Chapman}, S.~C., {Blain}, A.~W., {Alexander}, D.~M., {Ivison}, R.~J., 2005, \apj, 635, 853

\bibitem[{Canalizo \& Stockton(2000)}]{cana00}
{Canalizo}, G. \& {Stockton}, A., 2000, \apj, 528, 201

\bibitem[{Cisternas {et~al.}(2011){Cisternas},{Jahnke},{Inskip},{Kartaltepe},{Koekemoer},{Lisker},{Robaina},{Scodeggio},{Sheth},{Trump},{Andrae},{Miyaji},{Lusso},{Brusa},{Capak},{Cappelluti},{Civano},{Ilbert},{Impey},{Leauthaud},{Lilly},{Salvato},{Scoville},{Taniguchi}}]{cist11}
{Cisternas}, M., {Jahnke}, K., {Inskip}, K.~J., {Kartaltepe}, J., {Koekemoer}, A.~M., {Lisker}, T., {Robaina}, A.~R., {Scodeggio}, M., {Sheth}, K., {Trump}, J.~R., {Andrae}, R., {Miyaji}, T., {Lusso}, E., {Brusa}, M., {Capak}, P., {Cappelluti}, N., {Civano}, F., {Ilbert}, O., {Impey}, C.~D., {Leauthaud}, A., {Lilly}, S.~J., {Salvato}, M., {Scoville}, N.~Z., {Taniguchi}, Y., 2011, \apj, 726, 57

\bibitem[{Davies (2007)}]{dav07}
{Davies}, R.~I., 2007, \mnras, 375, 1099


\bibitem[{{Decarli} {et~al.}(2010){Decarli},{Falomo},{Treves},{Labita},{Kotilainen},{Scarpa}}]{deca10b}
{Decarli}, R., {Falomo}, R., {Treves}, A., {Labita}, M., {Kotilainen}, J.~K., {Scarpa}, R., 2010, \mnras, 402, 2453

\bibitem[{{Denney} {et~al.}(2009){Denney},{Peterson},{Dietrich},{Vestergaard},{Bentz}}]{denn08}
Denney, K.~D., Peterson, B.~M., Dietrich, M., Vestergaard, M., Bentz, M.~C., 2009, \apj, 692 246

\bibitem[{Eisenhauer} {et~al.}(2003)]{eis03}
Eisenhauer, F,. Abuter, R., Bickert, K., Biancat-Marchet, F., Bonnet, H., Brynnel, J., Conzelmann, R.~D., Delabre, B., Donaldson, R., Farinato, J., Fedrigo, E., Genzel, R., Hubin, N.~N., Iserlohe, C., Kasper, M.~E., Kissler-Patig, M., Monnet, G.~J., Roehrle, C., Schreiber, J., Stroebele, S., Tecza, M., Thatte, N.~A., Weisz, H., 2003, \procspie, 4841, 1548

\bibitem[{Emonts} {et~al.}(2008){Emonts},{Morganti},{Oosterloo},{Holt},{Tadhunter},{van der Hulst},{Ojha},{Sadler}]{emonts08}
{Emonts}, B.~H.~C., {Morganti}, R., {Oosterloo}, T.~A., {Holt}, J., {Tadhunter}, C.~N., {van der Hulst}, J.~M., {Ojha}, R., {Sadler}, E.~M., 2008, \mnras, 387, 197 

\bibitem[{Fathi {et~al.}(2005) Fathi, van de Ven, Peletier, Emsellem, Falc\'on-Barroso, Cappellari, de Zeeuw}]{fathi05}
Fathi, K., van de Ven, G., Peletier, R.\,F., Emsellem, E., Falc\'on-Barroso, J., Cappellari, M., de Zeeuw, T., 2005, \mnras, 364, 773

\bibitem[{Ferrarese \& Merrit(2000)}]{fm00}
Ferrarese, L., \& Merrit, D. 2000, \apj, 539, L9

\bibitem[{Gebhardt {et~al.}(2000)Gebhardt, Bender, Bower, Dressler, Faber,
  Filippenko, Green, Grillmair, Ho, Kormendy, Lauer, Magorrian, Pinkney,
  Richstone, \& Tremaine}]{geb00}
Gebhardt, K., Bender, R., Bower, G., Dressler, A., Faber, S.~M., Filippenko,
  A.~V., Green, R., Grillmair, C., Ho, L.~C., Kormendy, J., Lauer, T.~R.,
  Magorrian, J., Pinkney, J., Richstone, D., \& Tremaine, S. 2000, \apj, 539,
  L13

\bibitem[{Greene \& Ho (2005)}]{gh05}
{Greene}, J.~E., {Ho}, L.~C., 2005, \apj, 630, 122

\bibitem[{Graham (2004)}]{grah04}
{Graham}, A.~W., 2004, \apjl, 613, L33

\bibitem[{{Graham} {et~al.}(2011){Graham},{Onken},{Athanassoula},{Combes}}]{grah11}
{Graham}, A.~W., {Onken}, C.~A., {Athanassoula}, E., {Combes}, F., 2011, \mnras, 412, 2211

\bibitem[{{Greene} {et~al.}(2009){Greene},{Zakamska}, {Liu}, {Barth}, {Ho}}]{gre09}
{Greene}, J.~E.,  {Zakamska}, N.~L., {Liu}, X., {Barth}, A.~J., {Ho}, L.~C., 2009, \apj, 702, 441


\bibitem[{{Groves, Heckman \& Kauffmann} (2006){Groves}{Heckman}{Kauffmann}}]{groves06}
{Groves}, B.~A., {Heckman}, T.~M., {Kauffmann}, G., 2006, \mnras, 371, 1559


\bibitem[{{G{\"u}ltekin} {et~al.}(2009){G{\"u}ltekin}, {Richstone}, {Gebhardt},
  {Lauer}, {Tremaine}, {Aller}, {Bender}, {Dressler}, {Faber}, {Filippenko},
  {Green}, {Ho}, {Kormendy}, {Magorrian}, {Pinkney}, \& {Siopis}}]{g09}
{G{\"u}ltekin}, K., {Richstone}, D.~O., {Gebhardt}, K., {Lauer}, T.~R.,
  {Tremaine}, S., {Aller}, M.~C., {Bender}, R., {Dressler}, A., {Faber}, S.~M.,
  {Filippenko}, A.~V., {Green}, R., {Ho}, L.~C., {Kormendy}, J., {Magorrian},
  J., {Pinkney}, J., \& {Siopis}, C. 2009, \apj, 698, 198

\bibitem[{{H{\"a}ring} \& {Rix}(2004)}]{hr04}
{H{\"a}ring}, N., \& {Rix}, H.-W. 2004, \apj, 604, L89

\bibitem[{{Hirschmann} {et~al.}(2010){Hirschmann},{Khochfar},{Burkert},{Naab},{Genel},{Somerville}}]{hirs10}
Hirschmann, M., Khochfar, S., Burkert, A., Naab, T., Genel, S., Somerville, R.~S., 2010, \mnras, 407, 1016

\bibitem[{{Ho} {et~al.}(2008) {Ho, Darling \& Greene}}]{ho08}
{Ho}, L.~C., {Darling}, J., {Greene}, J.~E., 2008, \apj 681, 128

\bibitem[{{Ho} (2007)}]{ho07}
{Ho}, L.~C., 2007, \apj, 669, 821

\bibitem[{Hopkins} {et~al.}(2010){Hopkins}, {Bundy}, {Croton}, {Hernquist}, {Keres}, {Khochfar}, {Stewart}, {Wetzel}, {Younger}]{hopk10}
{Hopkins}, P.~F., {Bundy}, K., {Croton}, D., {Hernquist}, L., {Keres}, D., {Khochfar}, S., {Stewart}, K., {Wetzel}, A., {Younger}, J.~D., 2010, \apj, 715, 202

\bibitem[{{Hopkins \& Hernquist}(2009){Hopkins},{Hernquist}}]{hh09}
Hopkins, P.~F., Hernquist, L., 2009, \apj, 698, 1550

\bibitem[{Husemann} {et~al.}(2010){Husemann},{S\'{a}nchez},{Wisotzki},{Jahnke},{Kupko},{Nugroho},{Schramm}]{huse10}
{Husemann}, B., {S\'{a}nchez}, S.~F., {Wisotzki}, L., {Jahnke}, K., {Kupko}, D., {Nugroho}, D., {Schramm}, M., 2010, \aap, 519, 115 

\bibitem[{{Jahnke} \& {Maccio}(2011)}]{jahn11}
{Jahnke}, K., \& {Maccio}, A., 2011, \apjl, in press, arXiv:1006.0482

\bibitem[{{Jahnke} {et~al.}(2009){Jahnke}, {Bongiorno}, {Brusa}, {Capak},
  {Cappelluti}, {Cisternas}, {Civano}, {Colbert}, {Comastri}, {Elvis},
  {Hasinger}, {Ilbert}, {Impey}, {Inskip}, {Koekemoer}, {Lilly}, {Maier},
  {Merloni}, {Riechers}, {Salvato}, {Schinnerer}, {Scoville}, {Silverman},
  {Taniguchi}, {Trump}, \& {Yan}}]{jah09}
{Jahnke}, K., {Bongiorno}, A., {Brusa}, M., {Capak}, P., {Cappelluti}, N.,
  {Cisternas}, M., {Civano}, F., {Colbert}, J., {Comastri}, A., {Elvis}, M.,
  {Hasinger}, G., {Ilbert}, O., {Impey}, C., {Inskip}, K., {Koekemoer}, A.~M.,
  {Lilly}, S., {Maier}, C., {Merloni}, A., {Riechers}, D., {Salvato}, M.,
  {Schinnerer}, E., {Scoville}, N.~Z., {Silverman}, J., {Taniguchi}, Y.,
  {Trump}, J.~R., \& {Yan}, L. 2009, \apjl, 706, L215

\bibitem[{{Jahnke} {et~al.}(2004){Jahnke}, {Wisotzki}, {S\'anchez}, {Christensen}, {Becker}, {Kelz}, {Roth}}]{jahn04}
{Jahnke}, K., {Wisotzki}, L., {S\'anchez}, S.~F., {Christensen}, L., {Becker}, T., {Kelz}, A., {Roth} M.~M., 2004, Astronomische Nachrichten, 325, 128


\bibitem[{{Karim} {et~al.}(2011){Karim}, {Schinnerer}, {Martinez-Sansigre}, {Sargent}, {van der Wel}, {Rix}, {Ilbert}, {Smolcic}, {Carilli}, {Pannella}, {Koekemoer}, {Bell}, {Salvato}}]{karim}
{Karim}, A., {Schinnerer}, E., {Martinez-Sansigre}, A., {Sargent}, M.~T., {van der Wel}, A., {Rix}, H.-W., {Ilbert}, O., {Smolcic}, V., {Carilli}, C., {Pannella}, M., {Koekemoer}, A.~M., {Bell}, E.~F., {Salvato}, M., 2011, \apj, in press

\bibitem[{{Kelly} {et~al.}(2010){Kelly}, {Vestergaard}, {Fan}, {Hopkins}, {Hernquist}, {Siemiginowska}}]{kell10}
Kelly, B.~C., Vestergaard, M., Fan, X., Hopkins, P., Hernquist, L., Siemiginowska, A., 2010 \apj, 719, 1315


\bibitem[{Kennicutt}(1998)]{kenn98}
{Kennicutt}, R.~C., 1998, \araa, 36, 189


\bibitem[{{Kollmeier} {et~al.}(2006){Kollmeier}{Onken}{Kochanek}{Gould}{Weinberg}{Dietrich}{Cool}{Dey}{Eisenstein}{Jannuzi}{Le Floc'h}{Stern}}]{koll06}
{Kollmeier}, J.~A., {Onken}, C.~A., {Kochanek}, C.~S., {Gould}, A., {Weinberg}, D.~H., {Dietrich}, M., {Cool}, R., {Dey}, A., {Eisenstein}, D.~J., {Jannuzi}, B.~T., {Le Floc'h}, E., {Stern}, D., 2006, \apj, 648, 128

\bibitem[{{Kormendy} \& {Richstone}(1995)}]{kr95}
{Kormendy}, J., \& {Richstone}, D. 1995, \araa, 33, 581

\bibitem[{{Kormendy} \& {Bender}(2009)}]{kb09}
{Kormendy}, J., \& {Bender}, R. 2009, \apjl, 691, 142

\bibitem[{Magorrian {et~al.}(1998)Magorrian, Tremaine, Richstone, Bender,
  Bower, Dressler, Faber, Gebhardt, Green, Grillmair, Kormendy, \&
  Lauer}]{mag98}
Magorrian, J., Tremaine, S., Richstone, D., Bender, R., Bower, G., Dressler,
  A., Faber, S.~M., Gebhardt, K., Green, R., Grillmair, C., Kormendy, J., \&
  Lauer, T. 1998, \aj, 115, 2285

\bibitem[{{Marconi} \& {Hunt}(2003)}]{mh03}
{Marconi}, A., \& {Hunt}, L.~K. 2003, \apjl, 589, L21

\bibitem[{{Marconi} {et~al.}(2004){Marconi},{Risaliti}, {Gilli}, {Hunt}, {Maiolino}, {Salvati}}]{mar04}
Marconi, A., Risaliti, G., Gilli, R., Hunt, L.~K., Maiolino, R., Salvati, M., 2004, \mnras, 351, 169

\bibitem[{Markwardt} (2009)]{mark09}
Markwardt C., 2009, in Bohlender D., Dowler P., Durand D., eds, ASP Conf.
Ser. Vol. 411. Astronomical Data Analysis Software and Systems XVIII.
Astron. Soc. Pac., San Francisco, p. 251

\bibitem[{McGill {et~al.}(2008){McGill, Woo, Treu \& Malkan}}]{mcg08}
{McGill}, K.~L., {Woo}, J.-H., {Treu}, T., {Malkan}, M.~A., 2008, \apj, 673, 703


\bibitem[{McLure \& Dunlop(2002)}]{md02}
McLure, R.~J., \& Dunlop, J.~S. 2002, \mnras, 331, 795

\bibitem[{McLure \& Jarvis (2002)}]{mj02}
McLure, R.~J., \& Jarvis, M.~J., 2002, \mnras, 337, 109

\bibitem[{{Merloni} {et~al.}(2010){Merloni}, {Bongiorno}, {Bolzonella},
  {Brusa}, {Civano}, {Comastri}, {Elvis}, {Fiore}, {Gilli}, {Hao}, {Jahnke},
  {Koekemoer}, {Lusso}, {Mainieri}, {Mignoli}, {Miyaji}, {Renzini}, {Salvato},
  {Silverman}, {Trump}, {Vignali}, {Zamorani}, {Capak}, {Lilly}, {Sanders},
  {Taniguchi}, {Bardelli}, {Carollo}, {Caputi}, {Contini}, {Coppa}, {Cucciati},
  {de la Torre}, {de Ravel}, {Franzetti}, {Garilli}, {Hasinger}, {Impey},
  {Iovino}, {Iwasawa}, {Kampczyk}, {Kneib}, {Knobel}, {Kova{\v c}},
  {Lamareille}, {Le Borgne}, {Le Brun}, {Le F{\`e}vre}, {Maier}, {Pello},
  {Peng}, {Perez Montero}, {Ricciardelli}, {Scodeggio}, {Tanaka}, {Tasca},
  {Tresse}, {Vergani}, \& {Zucca}}]{mer10}
{Merloni}, A., {Bongiorno}, A., {Bolzonella}, M., {Brusa}, M., {Civano}, F.,
  {Comastri}, A., {Elvis}, M., {Fiore}, F., {Gilli}, R., {Hao}, H., {Jahnke},
  K., {Koekemoer}, A.~M., {Lusso}, E., {Mainieri}, V., {Mignoli}, M., {Miyaji},
  T., {Renzini}, A., {Salvato}, M., {Silverman}, J., {Trump}, J., {Vignali},
  C., {Zamorani}, G., {Capak}, P., {Lilly}, S.~J., {Sanders}, D., {Taniguchi},
  Y., {Bardelli}, S., {Carollo}, C.~M., {Caputi}, K., {Contini}, T., {Coppa},
  G., {Cucciati}, O., {de la Torre}, S., {de Ravel}, L., {Franzetti}, P.,
  {Garilli}, B., {Hasinger}, G., {Impey}, C., {Iovino}, A., {Iwasawa}, K.,
  {Kampczyk}, P., {Kneib}, J., {Knobel}, C., {Kova{\v c}}, K., {Lamareille},
  F., {Le Borgne}, J., {Le Brun}, V., {Le F{\`e}vre}, O., {Maier}, C., {Pello},
  R., {Peng}, Y., {Perez Montero}, E., {Ricciardelli}, E., {Scodeggio}, M.,
  {Tanaka}, M., {Tasca}, L.~A.~M., {Tresse}, L., {Vergani}, D., \& {Zucca}, E.
  2010, \apj, 708, 137

\bibitem[{More} (1978)]{more78}
More J., 1978, Numerical Analysis, G.~A. Watson, Springer-Verlag, Berlin, 630, 105

\bibitem[{Netzer \& Trakhtenbrot (2007)}]{nt07}
{Netzer}, H., {Trakhtenbrot}, B., 2007, \apj, 654, 754 


\bibitem[{Onken \& Kollmeier (2008)}]{ok08}
{Onken}, C.~A., {Kollmeier}, J.~A., 2008, \apj, 689, 13 


\bibitem[{{Onken}{et~al.}(2004){Onken}{Ferrarese}{Merritt}{Peterson}{Pogge}{Vestergaard}{Wandel}}]{onk04}
{Onken}, C.~A., {Ferrarese}, L., {Merritt}, D., {Peterson}, B.~M., {Pogge}, R.~W., {Vestergaard}, M., {Wandel}, A., 2004, \apj, 615, 645


\bibitem[{Oosterloo} { et~al.}(2010){Oosterloo},{Morganti},{Crocker},{J\"utte},{Cappellari},{de Zeeuw},{Krajnovi\'c},{McDermid},{Kuntschner},{Sarzi},{Weijmans}]{ost10}
{Oosterloo}, T.~A., {Morganti}, R.,  {Crocker}, A., {J\"utte}, E., {Cappellari}, M., {de Zeeuw}, T., {Krajnovi\'c}, D., {McDermid}, R., {Kuntschner}, H., {Sarzi}, M., {Weijmans}, A.-M., 2010, \mnras, 409, 500

\bibitem[{Oosterloo} { et~al.}(2007){Oosterloo},{Morganti},{Sadler},{van der Hulst},{Serra}]{ost07}
{Oosterloo}, T.~A., {Morganti}, R., {Sadler}, E.~M., {van der Hulst}, T., {Serra}, P., 2007, \aap, 465, 787

\bibitem[{{Peng}(2007)}]{peng07}
{Peng}, C.~Y., 2007, \apj, 671, 1098

\bibitem[{{Peng} {et~al.}(2006{\natexlab{a}}){Peng}, {Impey}, {Ho}, {Barton},
  \& {Rix}}]{peng06a}
{Peng}, C.~Y., {Impey}, C.~D., {Ho}, L.~C., {Barton}, E.~J., \& {Rix}, H.-W.
  2006{\natexlab{a}}, \apj, 640, 114

\bibitem[{{Peng} {et~al.}(2006{\natexlab{b}}){Peng}, {Impey}, {Rix},
  {Kochanek}, {Keeton}, {Falco}, {Leh{\'a}r}, \& {McLeod}}]{peng06b}
{Peng}, C.~Y., {Impey}, C.~D., {Rix}, H.-W., {Kochanek}, C.~S., {Keeton},
  C.~R., {Falco}, E.~E., {Leh{\'a}r}, J., \& {McLeod}, B.~A.
  2006{\natexlab{b}}, \apj, 649, 616

\bibitem[{{Persic} {et~al.}(1996){Persic, Salucci \& Stel}}]{per96}
{Persic}, M., {Salucci}, P., {Stel}, F., 1996, \mnras, 281, 27

\bibitem[{{Pickles}(1998)}]{pick98}
{Pickles}, A.~J., 1998, \pasp, 110, 863

\bibitem[{{Riechers} {et~al.}(2008){Riechers}, {Walter}, {Carilli}, {Bertoldi}, {Momjian}}]{riec08}
{Riechers}, D.~A., {Walter}, F., {Carilli}, C.~L., {Bertoldi}, F., {Momjian}, E., 2008, \apj, 686, L9

\bibitem[{{Riechers} {et~al.}(2009){Riechers}, {Walter}, {Carilli} \& {Lewis}}]{riec09}
{Riechers}, D.~A., {Walter}, F., {Carilli}, C.~L., {Lewis}, G.~F., 2009, \apj, 690, 463 

\bibitem[{{Robaina} {et~al.}(2010){Robaina}, {Bell}, {van der Wel}, {Somerville}, {Skelton}, {McIntosh}, {Meisenheimer}, {Wolf}}]{aday10}
Robaina, A.~R., Bell, E.~F., van der Wel, A., Somerville, R.~S., Skelton, R.~E., McIntosh, D.~H., Meisenheimer, K., Wolf, C., 2010, \apj, 719, 844

\bibitem[{{Salviander} {et~al.}(2007){Salviander}{Shields}{Gebhardt}{Bonning}}]{salv07}
{Salviander}, S.~B., {Shields}, G.~A., {Gebhardt}, K., {Bonning}, E.~W., 2007, \apj, 662, 131 

\bibitem[{{Sani} {et~al}(2011){Sani}{Marconi}{Hunt}{Risaliti}}]{sani}
{Sani}, E., {Marconi}, A., {Hunt}, L.~K., {Risaliti}, G., 2011, \mnras, 413, 1479

\bibitem[{{Sarzi} {et~al.}(2006){Sarzi},{Falc\'on-Barroso},{Davies},{Bacon},{Bureau},{Cappellari},{de Zeeuw},{Emsellem}, {Fathi},{Krajnovi\'c},{Kuntschner},{McDermid},{Peletier}}]{sarzi06}
{Sarzi}, M., {Falc\'on-Barroso}, J., {Davies}, R.~L., {Bacon}, R., {Bureau}, M., {Cappellari}, M., {de Zeeuw}, P.~T., {Emsellem}, E., {Fathi}, K., {Krajnovi\'c}, D., {Kuntschner}, H., {McDermid}, R.~M., {Peletier}, R.~F., 2006, \mnras, 366, 1151


\bibitem[{{Schneider} {et~al.}(2007) {Schneider}, {Hall}, {Richards}, {Strauss}, {Vanden Berk}, {Anderson}, {Brandt}, {Fan}, {Jester}, {Gray}, {Gunn}, {SubbaRao},  {Thakar},  {Stoughton},  {Szalay}, {Yanny}, {York},  {Bahcall},  {Barentine},  {Blanton},  {Brewington}, {Brinkmann}, {Brunner}, {Castander}, {Csabai},  {Frieman},  {Fukugita},  {Harvanek},  {Hogg},  {Ivezic}, {Kent}, {Kleinman}, {Knapp},  {Kron},  {Krzesinski}, {Long},  {Lupton},  {Nitta}, {Pier},  {Saxe},   {Shen}, {Snedden},  {Weinberg}, {Wu,}}]{schnei07}
Schneider, D.~P., Hall, P.~B., Richards, G.~T., Strauss, M.~A., Vanden Berk, D.~E., Anderson, S.~F., Brandt, W.~N., Fan, X., Jester, S., Gray, J., Gunn, J.~E., SubbaRao, M.~U., Thakar, A.~R., Stoughton, C., Szalay, A.~S., Yanny, B., York, D.~G., Bahcall, N.~A., Barentine, J., Blanton, M.~R., Brewington, H., Brinkmann, J., Brunner, R.~J., Castander, F.~J., Csabai, I., Frieman, J.~A., Fukugita, M., Harvanek, M., Hogg, D.~W., Ivezic, Z., Kent, S.~M., Kleinman, S.~J., Knapp, G.~R., Kron, R.~G., Krzesinski, J., Long, D.~C., Lupton, R.~H., Nitta, A., Pier, J.~R., Saxe, D.~H.,  Shen, Y., Snedden, S.~A., Weinberg, D.~H., Wu, J., 2007, \aj, 134, 102

\bibitem[{{Shen} {et~al.}(2008){Shen}, {Greene}, {Strauss}, {Richards}, {Schneider}}]{shen08}
Shen, Y., Greene, J.~E., Strauss, M.~A., Richards, G.~T., Schneider, D.~P., 2008, \apj, 680, 169

\bibitem[{Shen \& Kelly (2010)}]{sk10}
Shen, Y., Kelly, B.~C., 2010, \apj, 713, 41

\bibitem[{{Shields} {et~al.}(2003){Shields}, {Gebhardt}, {Salviander}, {Wills}, {Xie}, {Brotherton}, {Yuan}, {Dietrich}}]{shie03} 
{Shields}, G.~A., {Gebhardt}, K., {Salviander}, S., {Wills}, B.~J., {Xie}, B., {Brotherton}, M.~S., {Yuan}, J., {Dietrich}, M., 2003, \apj, 583, 124

\bibitem[{{Shields} {et~al.}(2006){Shields}, {Menezes}, {Masart} \&  {Vanden Bout}}]{shie06} 
{Shields}, G.~A., {Menezes}, K.~L., {Masart}, C.~A., {Vanden Bout}, P., 2006, \apj, 641, 683

\bibitem[{{Somerville} (2009) {Somerville}}]{som09}
{Somerville}, R.~S., 2009, \mnras, 399, 1988

\bibitem[{{Tremaine} {et~al.}(2002){Tremaine}, {Gebhardt}, {Bender}, {Bower},
  {Dressler}, {Faber}, {Filippenko}, {Green}, {Grillmair}, {Ho}, {Kormendy},
  {Lauer}, {Magorrian}, {Pinkney}, \& {Richstone}}]{tre02}
{Tremaine}, S., {Gebhardt}, K., {Bender}, R., {Bower}, G., {Dressler}, A.,
  {Faber}, S.~M., {Filippenko}, A.~V., {Green}, R., {Grillmair}, C., {Ho},
  L.~C., {Kormendy}, J., {Lauer}, T.~R., {Magorrian}, J., {Pinkney}, J., \&
  {Richstone}, D. 2002, \apj, 574, 740%

\bibitem[{{Treu} {et~al.}(2007){Treu}, {Woo}, {Malkan}, \&
  {Blandford}}]{treu07}
{Treu}, T., {Woo}, J., {Malkan}, M.~A., \& {Blandford}, R.~D. 2007, \apj, 667,
  117

\bibitem[{{Yu \& Tremaine} (2002){Yu}, {Tremaine}}]{yt02}
Yu, Q., Tremaine, S., 2002, \mnras, 335, 965

\bibitem[{{van de Ven \& Fathi} (2010){van de Ven} \&
  {Fathi}}]{vdven2010}
{van de Ven}, G., \& {Fathi}, K., 2010, \apj, 723,
  767

\bibitem[{{van Dokkum} {et~al.}(2010){van Dokkum}, {Whitaker}, {Brammer}, {Franx}, {Kriek}, {Labb\'e}, {Marchesini}, {Quadri}, {Bezanson}, {Illingworth}, {Muzzin}, {Rudnick}, {Tal}, {Wake}}]{vd10}
{van Dokkum}, P.~G., {Whitaker}, K.~E., {Brammer}, G., {Franx}, M., {Kriek}, M., {Labb\'e}, I., {Marchesini}, D., {Quadri}, R., {Bezanson}, R., {Illingworth}, G.~D., {Muzzin}, A., {Rudnick}, G., {Tal}, T., {Wake}, D., 2010, \apj, 709, 1018

\bibitem[{{Veilleux \& Osterbrock} (1987){Veilleux}{Osterbrock}}]{vo}
{Veilleux}, S., {Osterbrock}, D.~E., 1987, \apjs, 63, 295

\bibitem[{{Vestergaard} (2002) {Vestergaard}}]{vest02}
{Vestergaard}, M., 2002, \apj, 571, 733

\bibitem[{{Vestergaard \& Peterson} (2006){Vestergaard}, {Peterson}}]{vp06}
{Vestergaard}, M., {Peterson}, B.~M., 2006, \apj, 641, 689 

\bibitem[{{Walter} {et~al.}(2004){Walter}, {Carilli}, {Bertoldi}, {Menten},
  {Cox}, {Lo}, {Fan}, \& {Strauss}}]{wal04}
{Walter}, F., {Carilli}, C., {Bertoldi}, F., {Menten}, K., {Cox}, P., {Lo},
  K.~Y., {Fan}, X., \& {Strauss}, M.~A. 2004, \apj, 615, L17

\bibitem[{{White} {et~al.}(1997){White},{Becker},{Helfand},{Gregg}}]{first2}
{White}, R.~L., {Becker}, R.~H., {Helfand}, D.~J., {Gregg}, M.~D., 1997, \apj, 475, 479


\bibitem[{{Woo} {et~al.}(2006){Woo}, {Treu}, {Malkan}, \& {Blandford}}]{woo06}
{Woo}, J., {Treu}, T., {Malkan}, M.~A., \& {Blandford}, R.~D. 2006, \apj, 645,
  900



\end{thebibliography}
\end{document}